\documentclass[12pt]{article}
\usepackage{putex}

\makeatletter
\DeclareRobustCommand*{\bfseries}{%
  \not@math@alphabet\bfseries\mathbf
  \fontseries\bfdefault\selectfont
  \boldmath
}
\makeatother

\input{glyphtounicode}
\def\Res{\mathop{\rm Res}}
\def\deltaeq{\stackrel{\delta}{=}}

\begin{document}

\preprint{PUPT-2519}

\title{$O(N)$ and $O(N)$ and $O(N)$}
\authors{Steven S. Gubser, Christian Jepsen, Sarthak Parikh, and Brian Trundy}
\institution{PU}{Joseph Henry Laboratories, Princeton University, Princeton, NJ 08544, USA}

\abstract{Three related analyses of $\phi^4$ theory with $O(N)$ symmetry are presented.  In the first, we review the $O(N)$ model over the $p$-adic numbers and the discrete renormalization group transformations which can be understood as spin blocking in an ultrametric context.  We demonstrate the existence of a Wilson-Fisher fixed point using an $\epsilon$ expansion, and we show how to obtain leading order results for the anomalous dimensions of low dimension operators near the fixed point.  Along the way, we note an important aspect of ultrametric field theories, which is a non-renormalization theorem for kinetic terms.  In the second analysis, we employ large $N$ methods to establish formulas for anomalous dimensions which are valid equally for field theories over the $p$-adic numbers and field theories on $\mathbb{R}^n$.  Results for anomalous dimensions agree between the first and second analyses when they can be meaningfully compared.  In the third analysis, we consider higher derivative versions of the $O(N)$ model on $\mathbb{R}^n$, the simplest of which has been  studied in connection with spatially modulated phases.  Our general formula for anomalous dimensions can still be applied.  Analogies with two-derivative theories hint at the existence of some interesting unconventional field theories in four real Euclidean dimensions.}

\date{March 2017}
\maketitle

\tableofcontents

\section{Introduction}
\label{INTRODUCTION}
Early literature on the renormalization group, notably Wilson's papers \cite{Wilson:1971bg,Wilson:1971dh} and the classic review by Wilson and Kogut \cite{Wilson:1973jj}, makes prominent use of finite-step recursion relations to approximate renormalization group flow in a continuum field theory.  The recursion relations are in the spirit of Kadanoff's block-spin version of the renormalization group \cite{Kadanoff:1966wm}, and it is noted in Wilson's early work that the recursion relations become exact when used on the Dyson hierarchical model \cite{Dyson:1968up}, while when applied to ordinary $\phi^4$ field theory on $\mathbb{R}^n$ they can be used to extract scaling dimensions at the Wilson-Fisher fixed point that are correct through order $\epsilon$, where $\epsilon = 4-n$.

Subsequent work, including \cite{bleher1973,bleher1975} and reviewed in \cite{bleher1987}, established rigorous results on the solvability and fixed points of the renormalization group for hierarchical models as realized by the recursion relations.  In \cite{Lerner:1989ty} it was understood that a continuum limit of a suitable hierarchical model gives $p$-adic $\phi^4$ theory---meaning a theory with a real field $\phi$ which is a function defined over the $p$-adic numbers $\mathbb{Q}_p$.  One may similarly treat theories where $\mathbb{Q}_p$ is replaced by a field extension $\mathbb{Q}_{p^n}$, which in part means working with a vector space $\mathbb{Q}_p^n$.\footnote{Dyson's original approach, in which spin variables are grouped in pairs, and then the pairs are paired, etc., gives rise in the continuum limit to a field theory over $\mathbb{Q}_2$, whereas if $q = p^n$ spins are grouped together at each step one gets in an appropriate limit a field theory over $\mathbb{Q}_{p^n}$.  The spirit of this construction does not seem to require that $q=p^n$ is a power of a prime.  However, if it is not, ``$\mathbb{Q}_q$'' is not a field, nor even an integral domain, and it is harder to understand either the $q$-adic norm which enters into correlators or the $q$-adic conformal symmetry that arises near a critical point.  We therefore leave the interesting point of general composite $q$ to future work.} The $O(N)$ generalization of \cite{Lerner:1989ty} was first considered in \cite{Okada:1988hv}, which however restricted the form of the kinetic term and did not consider extensions of the $p$-adic field, and for these reasons dealt only with an asymptotically free theory. These restrictions were lifted in \cite{Missarov:2006in,Missarov:2006iu,missarov2012p}, in which a renormalized projection Hamiltonian formalism was used to explore the Wilson-Fisher fixed point of the $p$-adic $O(N)$ model in an appropriate $\epsilon$ expansion (similar to expanding in $\epsilon=4-n$ dimensions in standard $O(N)$ field theory on $\mathbb{R}^n$) and to compute the critical exponents. In section~\ref{WILSONIAN} we review the Wilsonian renormalization of the $p$-adic $O(N)$ model and carry out standard diagrammatic perturbation theory in the $\epsilon$ expansion.\footnote{Although we do not pursue holographic calculations in the current work, it is natural to hope that the $p$-adic $O(N)$ model for large $N$ is dual to some appropriate modification of Vasiliev theory defined on the Bruhat-Tits tree, along the lines of \cite{Klebanov:2002ja,Giombi:2009wh,Gubser:2016guj,Heydeman:2016ldy,Gubser:2016htz}.}

In section~\ref{VASILIEV}, we adapt methods of \cite{Vasiliev:1981yc,Vasiliev:1981dg} to the $p$-adic case to obtain self-consistent results for the critical exponents of the non-Gaussian fixed point that are exact in $\epsilon$ and valid through the first non-trivial order in large $N$.  We demonstrate agreement between these two approaches where they overlap, namely small $\epsilon$ and large $N$.  The large $N$ methods are based mostly on position space integrals of multiplicative characters, and by defining suitable variants of the Euler beta function motivated by the simplest of these integrals we are able to give universal formulas for the anomalous dimensions in terms of residues at poles of meromorphic functions which are simple rational combinations of beta functions.

Readers wishing to skip technical details and see in section~\ref{HIGHER} the final results for anomalous dimensions should be aware of some non-standard notation in this paper: $\zeta(t)$ is not the usual Riemann zeta function, nor are $\Gamma(t)$ and ${\rm B}(t_1,t_2)$ the usual gamma or beta functions.  Instead $\zeta(t)$ is a ``local'' version of the zeta function defined either for $\mathbb{R}$ or for $\mathbb{Q}_p$, while $\Gamma(t)$ and ${\rm B}(t_1,t_2)$ are defined in reference either to $\mathbb{R}^n$ or the unramified extension $\mathbb{Q}_{p^n}$ of the $p$-adic numbers of degree $n$.  All these functions take complex arguments and are meromorphic.  The point of defining $\zeta$, $\Gamma$, and ${\rm B}$ anew every time we pass to a new field or vector space is that physical quantities like scaling dimensions tend to have a universal form when expressed in terms of the appropriate functions.  Even well-known results in $\mathbb{R}^n$ assume pleasingly simple forms in terms of suitably defined $\Gamma_{\mathbb{R}^n}(t)$ and ${\rm B}_{\mathbb{R}^n}(t_1,t_2)$.

Formulas applicable only to $\mathbb{R}^n$ or only to $\mathbb{Q}_{p^n}$ will be suitably marked: For example,
 \eqnRn{zetaDefRn}{
  \zeta(t) = \zeta_{\mathbb{R}}(t) \equiv \pi^{-t/2} \Gamma_{\rm Euler}(t/2)
 }
whereas
 \eqnQq{zetaDefQpn}{
  \zeta(t) = \zeta_{\mathbb{Q}_p}(t) \equiv {1 \over 1 - p^{-t}} \,.
 }
Note that the volume of $S^{n-1}$ is $2/\zeta_{\mathbb{R}}(n)$, whereas $1/\zeta_{\mathbb{Q}_p}(n)$ is the volume of the set of units in $\mathbb{Q}_{p^n}$, which is the set of elements $\xi \in \mathbb{Q}_{p^n}$ with $|\xi|=1$.  It is tempting to define $\zeta_{\mathbb{R}^n}(t) \equiv \zeta_{\mathbb{R}}(nt)$ and $\zeta_{\mathbb{Q}_{p^n}}(t) \equiv \zeta_{\mathbb{Q}_p}(nt)$, along the lines of \cite{Gelfand:1969a,Brekke:1993gf}, but for our current purposes it is clearer not to do so, and instead always to construe $\zeta(t)$ as $\zeta_{\mathbb{R}}(t)$ or $\zeta_{\mathbb{Q}_p}(t)$. 

Formulas which apply equally to $\mathbb{R}^n$ and $\mathbb{Q}_{p^n}$ will be left unmarked: For example,\footnote{Gamma functions defined this way satisfy the functional equation $\Gamma(t) \Gamma(n-t) = 1$. This results in the following useful identity for the beta function: ${\rm B}(t_1, t_2) = {\rm B}(t_1, n-t_1-t_2) = {\rm B}(t_2, n-t_1-t_2)$.}
 \eqn{GammaBeta}{
  \Gamma(t) \equiv {\zeta(t) \over \zeta(n-t)} \qquad\qquad
  {\rm B}(t_1,t_2) \equiv {\Gamma(t_1) \Gamma(t_2) \over \Gamma(t_1+t_2)} \,.
 }
In the same spirit as $\Gamma$ and $\rm B$ implicitly refer either to $\mathbb{R}^n$ or to $\mathbb{Q}_{p^n}$, we also use $|x|$ to denote the absolute value in either $\mathbb{R}^n$ or $\mathbb{Q}_{p^n}$.  In the former case, $|x| = \sqrt{\sum_{i=1}^n x_i^2}$, which is an Archimedean norm.  In the latter case, $|x|$ is ultrametric and takes values which are integer powers of $p$; formally, $|x|$ is defined as the $p$-adic norm of the field norm of $x$ with respect to the extension relation $\mathbb{Q}_{p^n}:\mathbb{Q}_p$.\footnote{An introduction to $\mathbb{Q}_{p^n}$ and other concepts used in the current work can be found in \cite{Gubser:2016guj}, and closely related ideas have appeared in \cite{Heydeman:2016ldy}.  Here let us note that the distinction between ultrametric and Archimedean norms hinges on the triangle inequality.  An ultrametric norm must satisfy $|x+y| \leq \max\{|x|,|y|\}$, which implies the triangle inequality but is obviously stronger than it.  An Archimedean norm satisfies the standard triangle inequality and also has the property that if $0 < |x| < |y|$, then there exists some integer $n$ such that $|nx| > |y|$, where $nx$ is understood as $x$ added to itself $n$ times.  The ultrametric and Archimedean properties are mutually exclusive.  Ostrowski's Theorem states (approximately) that the only norms on the rational numbers are the usual Archimedean norm together with the $p$-adic norms for any prime $p$.}

A key feature of ultrametric theories is that their kinetic terms are non-local.  In momentum space, they are expressed as $\int dk \, {1 \over 2} \phi(-k) |k|^s \phi(k)$, where $s$ is a spectral parameter, a real number which we must usually choose between $0$ and $n$.  This makes ultrametric theories similar to bilocal field theories on $\mathbb{R}^n$ as studied in \cite{Fisher:1972zz,Sak:1973a} and more recently, for example, in \cite{Paulos:2015jfa}.  In these bilocal theories, similar kinetic terms are considered, with $|k|^s$ as their momentum space kernel.  A special feature of field theories on $\mathbb{R}^n$ is that when $s$ is a positive even integer, the kinetic term becomes local in position space.  In section~\ref{VASILIEV}, we mostly focus on the case $s=2$ when we examine field theories on $\mathbb{R}^n$.  In section~\ref{HIGHER}, we argue that $s=4$ and higher even integers are also interesting: these values give rise to higher derivative $O(N)$ models, and they seem to be free of pathologies as long as they are regarded as Euclidean path integral field theories. An example of such a theory was discussed already in \cite{Gava:1978gd}. Four derivative theories have also been considered in the condensed matter literature \cite{Hornreich:1975a}, where they have been used to investigate spatially modulated phases \cite{Michelson:1977a} along the lines of the Landau-Brazovskii model \cite{Brazovskii:1975a}.  Commonly called Lifshitz points, these four derivative theories have connections with the next-to-nearest neighbor Ising model, as reviewed in \cite{Selke:1988a}.  (The main focus of many of the condensed matter applications is anisotropic models, in which one direction is singled out and may exhibit different scaling behavior.  In the current work, we are instead interested in the isotropic case.)  We will recall the basics of these approaches in section~\ref{HIGHER}.

\section{$p$-adic Wilsonian renormalization}
\label{WILSONIAN}

Let $\phi^i$ be a map from $\mathbb{Q}_{p^n}$ to $\mathbb{R}^N$, where $n$ and $N$ are positive integers and $p$ is a prime number.  (If $n=1$ then the domain of $\phi^i$ is the $p$-adic numbers themselves).  Our reason for focusing on the unramified extension $\mathbb{Q}_{p^n}$ is that it is an $n$-dimensional vector space over $\mathbb{Q}_p$ with a natural ultrametric norm taking the same values as the norm on $\mathbb{Q}_p$, similar to the way $\mathbb{R}^n$ is an $n$-dimensional vector space over $\mathbb{R}$ with a natural Archimedean norm (namely the usual $L^2$ norm).  It is likely that the discussion to follow could be generalized to somewhat more general ultrametric spaces, but we do not pursue this.

\subsection{Action}
Following \cite{Lerner:1989ty}, we consider the action
 \eqnQq{ONmodel}{
  S &= \int dk \, {1 \over 2} \phi^i(-k) (|k|^s + r) \phi^i(k)  \cr & \;\; {}+ 
   \int dk_1 dk_2 dk_3 dk_4 \, \delta(k_1+k_2+k_3+k_4) {\lambda \over 4!} T_{i_1i_2i_3i_4} 
    \phi^{i_1}(k_1) \phi^{i_2}(k_2) \phi^{i_3}(k_3) \phi^{i_3}(k_4) \,,
 }
where summation over repeated indices is implied, and following \cite{Kleinert:2001ax} we set
 \eqnQq{TDef}{
  T_{i_1i_2i_3i_4} = {1 \over 3} (\delta_{i_1i_2} \delta_{i_3i_4} + 
    \delta_{i_1i_3} \delta_{i_2i_4} + \delta_{i_1i_4} \delta_{i_2i_3}) \,.
 }
Fourier transforms are defined by
 \eqnQq{FourierDef}{
  \phi^i(x) = \int dk \, \chi(kx) \phi^i(k) \,,
 }
where $\chi(\xi) = e^{2\pi {\rm i} \{ \xi \}}$ is an additive character on $\mathbb{Q}_{p^n}$.  All integrals in \eno{ONmodel} and \eno{FourierDef} are by default over all of $\mathbb{Q}_{p^n}$; however, we may impose a hard momentum cutoff $|k| \leq \Lambda$ where $\Lambda$ is an integer power of $p$ and $|k|$ is the standard norm on $\mathbb{Q}_{p^n}$, whose values are integer powers of $p$.

The $O(N)$ model on $\mathbb{Q}_{p^n}$ comes with three real parameters, $r$ (a mass-squared parameter), $\lambda$, and a spectral parameter $s$ which tells us in the free theory that the dimension of $\phi(x)$ is ${n-s \over 2}$.  Unlike in ordinary local field theories on $\mathbb{R}^n$, $s$ is an adjustable parameter in a $p$-adic context.

\subsection{One-loop amplitudes}

\begin{figure}[h!]
		\center
			\includegraphics{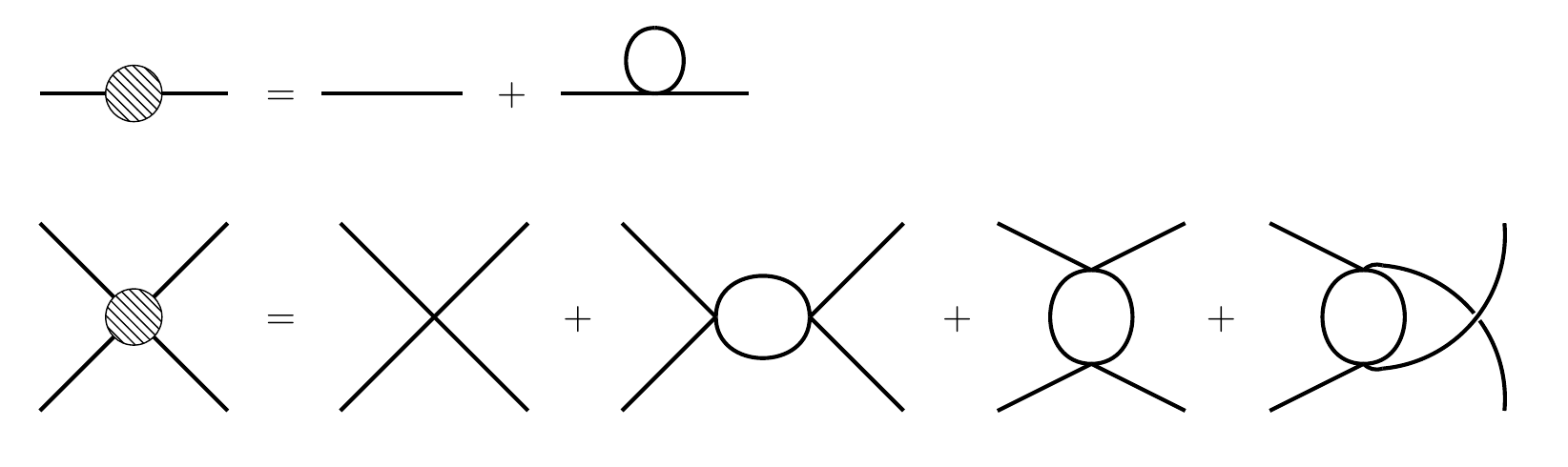}
			\caption{Diagrammatic representation of the two- and four-point functions to one loop order.}			\label{TwoFourGraphs}
		\end{figure}

To renormalize $\phi^4$ theory we typically need to handle divergences in the two-point and four-point functions.  To one loop order, these Green's functions take the following forms:
 \eqnQq{TwoFourGreen}{
  G^{(2)}_{ij}(k) &= {\delta_{ij} \over |k|^s + r} + 
    {\delta_{ij} \over (|k|^s+r)^2} {1 \over 2} (-\lambda) {N+2 \over 3} I_2
   = {\delta_{ij} \over |k|^s + r + \lambda {N+2 \over 6} I_2} + {\cal O}(\lambda^2) \cr
  G^{(4)}_{i_1i_2i_3i_4}(k_i) &= -\lambda T_{i_1i_2i_3i_4} + {1 \over 2} (-\lambda)^2 
    {N+8 \over 9} (I_4^{(s)} + I_4^{(t)} + I_4^{(u)}) T_{i_1i_2i_3i_4} \,,
 }
In \eno{TwoFourGreen} and below, we omit the momentum-conserving delta functions from the Green's functions.  The loop integrals are
 \eqnQq{LoopIntegrals}{
  I_2 = \int {d\ell \over |\ell|^s + r} \qquad\qquad
  I_4^{(S)} = \int {d\ell \over (|\ell|^s + r)(|\ell+k_1+k_2|^s + r)} \,.
 }
$I_4^{(T)}$ and $I_4^{(U)}$ are defined like $I_4^{(S)}$, but with $k_1+k_2$ replaced by $k_1+k_3$ for $I_4^{(T)}$ and by $k_1+k_4$ for $I_4^{(U)}$. A diagrammatic account of the formulas \eno{TwoFourGreen} is summarized in figure~\ref{TwoFourGraphs}.  The standard challenge of perturbative renormalization group analysis is to tame divergences at large $|\ell|$ (the ultraviolet) arising in the integrals \eno{LoopIntegrals}.

\subsection{Wilsonian renormalization}
\label{RENORMALIZATION}

In a Wilsonian approach, we integrate out a shell of hard momenta, so we want the internal momenta, denoted $\ell$ in \eno{TwoFourGreen}-\eno{LoopIntegrals}, to be hard, while the external momenta $k_i$ are soft.  A key property of $\mathbb{Q}_{p^n}$ is that it organizes into momentum shells whose magnitudes are integer powers of $p$, and we can integrate out one such momentum shell at a time.  Momentum shell integration is easy to do because the integrands are constant over each momentum shell.  Explicitly,
 \eqnQq{ItwoUltrametric}{
  I_2 &= \int_{|\ell|=\Lambda} {d\ell \over \Lambda^s + r} = {1 \over \zeta(n)}
    {\Lambda^n \over \Lambda^s + r}  \cr
  I_4 &= \int_{|\ell|=\Lambda} {d\ell \over (\Lambda^s + r)^2} = {1 \over \zeta(n)}
    {\Lambda^n \over (\Lambda^s + r)^2} \,.
 }
The result for $I_4$ is the same for all three channels (so we dropped the channel label), and it relies on the fact that $|\ell+k| = |\ell|$ when $\ell$ is hard and $k$ is soft.  This equality is an exact statement which follows directly from $|k| < |\ell|$ together with the ultrametric property of the norm on $\mathbb{Q}_{p^n}$.  The situation contrasts strongly with the Archimedean case, where we have the weaker condition $|\ell+k| \approx |\ell|$ when $|k| \ll |\ell|$.

To extract the recursion relations that define the renormalization group for the $p$-adic $O(N)$ model, we require that $G^{(2)}_{ij}(k)$ as computed in \eno{TwoFourGreen} through one-loop order, with the loop momentum required to satisfy $|\ell| = \Lambda$, should coincide with the tree level Green's function $G^{(2)}_{{\rm soft},ij}(k) = {\delta_{ij} \over |k|^s + r_{\rm soft}}$ of an effective soft theory with a hard momentum cutoff at $\Lambda/p$ instead of $\Lambda$.  Likewise, we seek to have $G^{(4)}_{i_1i_2i_3i_4}$ as computed in \eno{TwoFourGreen} coincide with the tree-level $G^{(4)}_{{\rm soft},i_1i_2i_3i_4} = -\lambda_{\rm soft} T_{i_1i_2i_3i_4}$, and this is possible because $G^{(4)}_{i_1i_2i_3i_4}$ has no momentum dependence (beyond the momentum-conserving delta function which we have suppressed).  Altogether, we find
 \eqnQq{rlambda}{
  r_{\rm soft} &= r + \lambda {N+2 \over 6} {1 \over \zeta(n)} 
    {\Lambda^n \over \Lambda^s + r}  \cr
  \lambda_{\rm soft} &= \lambda - \lambda^2 {N+8 \over 6} {1 \over \zeta(n)} 
    {\Lambda^n \over (\Lambda^s + r)^2} \,.
 }
The relations \eno{rlambda} are more simply expressed in terms of analogs of ``dimensionless couplings''
 \eqnQq[c]{rlambdaBar}{
  \bar{r} = {r \over \Lambda^{[r]}} \qquad\qquad
  \bar{r}_{\rm soft} = {r_{\rm soft} \over (\Lambda/p)^{[r]}}  \cr
  \bar\lambda = {\lambda \over \Lambda^{[\lambda]}} \qquad\qquad
  \bar\lambda_{\rm soft} = {\lambda_{\rm soft} \over (\Lambda/p)^{[\lambda]}} \,,
 }
where
 \eqnQq{FreeDims}{
  [r] = s \qquad\qquad [\lambda] = \epsilon \equiv 2s - n \,.
 }
In general, $[X]$ is the dimension of a quantity $X$ for the Gaussian fixed point at $\lambda=0$.  Thus for example $[|k|] = 1$, $[dk] = n$, $[|x|] = -1$, and $[\phi(x)] = {n-s \over 2}$.  Holding $n$ fixed and increasing $s$ is analogous to holding $s$ fixed and lowering $n$, and the analog of the upper critical dimension, at which $\phi^4$ becomes marginal, is $s = n/2$.\footnote{The analog of the lower critical dimension is $s=n$, so IR critical behavior occurs for  $n/2<s<n$, or equivalently $s< n< 2s$ which is the analog of $2<n<4$ in $\phi^4$ theory in $\mathbb{R}^n$.}  Thus $[\lambda]$ itself is the analog of the parameter $\epsilon = 4-n$ in ordinary $\phi^4$ theory, and in some formulas we emphasize this by writing quantities in terms of $\epsilon$.

Having defined dimensionless couplings in \eno{rlambdaBar}, we can now recast \eno{rlambda} as
 \eqnQq{rlambdaRG}{
  \bar{r}_{\rm soft} &= p^s \left[ \bar{r} + \bar\lambda {N+2 \over 6} {1 \over \zeta(n)}
    {1 \over 1 + \bar{r}} \right]  \cr
  \bar\lambda_{\rm soft} &= p^\epsilon \left[ \bar\lambda - 
    \bar\lambda^2 {N+8 \over 6} {1 \over \zeta(n)} {1 \over (1+\bar{r})^2} 
    \right] \,.
 }
These are the recursion relations which define the renormalization of the $p$-adic $O(N)$ model through one loop.

\subsection{A non-renormalization theorem}
\label{NONRENORMALIZATION}

\begin{figure}[h!]
		\center
			\includegraphics{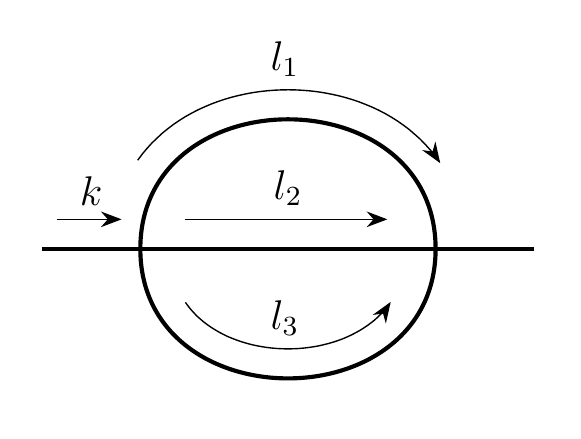}
			\caption{The underground diagram, the lowest order diagram that contributes to wave--function renormalization in Archimedean $\phi^4$ theory.}			\label{Underground}
		\end{figure}

Note that we didn't have to worry about wave-function renormalization when working out the recursion relations \eno{rlambdaRG}.  Absence of wave-function renormalization is a trivial observation at this loop order, since there is no way to get momentum dependence in the one-loop correction to $G^{(2)}_{ij}(k)$ even in an Archimedean theory.  A striking point about the $p$-adic $O(N)$ model is that (at least in a perturbative Wilsonian approach), no wave-function renormalization ever occurs.  Better yet, no diagrammatic loop correction {\it ever} exhibits momentum dependence, even in higher point amplitudes.  That is, the effective action is always schematically of the form\footnote{By $\int dx \, V_{\rm eff}(\vec\phi(x))$ we really mean a sum of powers of $\vec\phi(x)$, suitably contracted with $O(N)$-covariant tensors and multiplied by running couplings like $r$ and $\lambda$, and expressed in momentum space as integrals against momentum-conserving delta functions.}
 \eqnQq{GeneralEffective}{
  S = \int dk \, {1 \over 2} \vec\phi(-k) |k|^s \cdot \vec\phi(k) +
    \int dx \, V_{\rm eff}(\vec\phi(x)) \,,
 }
where $V_{\rm eff}(\vec\phi(x))$ undergoes renormalization group flow but the ``kinetic term'' $\phi(-k) |k|^s \phi(k)$ is never renormalized, nor are any other $k$-dependent terms generated as they are for theories on $\mathbb{R}^n$.  In other words, the renormalization group acts strictly on the purely non-derivative, local part of the action which depends on $\vec\phi(x)$ at one point only.  This feature of the renormalization group seems to have been appreciated already for the hierarchical model \cite{bleher1987}.  It hinges on ultrametricity, as we can see by examining the first diagram whose momentum dependence would ordinarily contribute to wave-function renormalization in $\phi^4$ theory, namely the underground diagram shown in figure~\ref{Underground}.  The loop integral is
 \eqnQq{Iunderground}{
  I_{2'} = \int_{|\ell_1|=\Lambda} d\ell_1 \int_{|\ell_2|=\Lambda} d\ell_2 
    \int_{|\ell_3|=\Lambda} d\ell_3 \, {\delta(\ell_1+\ell_2+\ell_3-k) \over
    (\Lambda^s + r)^3}\,.
 }
To see that $I_{2'}$ is actually independent of $k$, we use the $u$-substitution $\tilde\ell_3 = \ell_3 - k$.  Ultrametricity guarantees that the map $\ell_3 \to \tilde\ell_3$ is a bijection from the momentum shell $|\ell_3| = \Lambda$ to itself, provided $|k| < \Lambda$. Similar arguments can be applied to general Feynman diagrams \cite{Lerner:1989ty}. 

\subsection{Fixed point and anomalous dimensions}

Finding a fixed point of the discrete RG equations \eno{rlambdaRG} now amounts to setting $\bar{r}_{\rm soft} = \bar{r}$ and $\bar\lambda_{\rm soft} = \bar\lambda$.  This happens, to leading order in small $\epsilon$, at the $p$-adic Wilson-Fisher fixed point,
 \eqnQq{WF}{
  \bar{r}_* = -\zeta(n/2) {N+2 \over N+8} \epsilon \log p \qquad\qquad
  \bar\lambda_* = {6 \zeta(n) \over N+8} \epsilon \log p \,.
 }
To analyze anomalous dimensions at the fixed point, we consider perturbations
 \eqnQq{PerturbWF}{
  \bar{r} = \bar{r}_* + \delta\bar{r} \qquad\qquad
  \bar\lambda = \bar\lambda_* + \delta\bar\lambda \,.
 }
To linear order in $\delta\bar{r}$ and $\delta\bar\lambda$, the discrete RG equations become
 \eqnQq{LinearizedRG}{
  \begin{pmatrix} \delta\bar{r}_{\rm soft} \\ \delta\bar\lambda_{\rm soft} \end{pmatrix} = 
    M \begin{pmatrix} \delta\bar{r} \\ \delta\bar\lambda \end{pmatrix} \,.
 }
The explicit form of $M$ can be worked out easily starting from \eno{rlambdaRG} but is unenlightening.  Eigenvalues of $M$ take the form $p^{n-\Delta}$ where $\Delta$ is the dimension of a primary operator ${\cal O}$ in the fixed point theory.  To see this, note that if $\rho$ is the coupling dual to ${\cal O}$, then $\rho$ has dimension $n-\Delta$, and we naturally define $\bar\rho = \rho/\Lambda^{n-\Delta}$, while $\bar\rho_{\rm soft} = \rho/(\Lambda/p)^{n-\Delta} = p^{n-\Delta} \bar\rho$.   By straightforward calculation, we see that the dimensions from \eno{LinearizedRG} to leading order in small $\epsilon$ take the form
 \eqnQq{TwoDimensions}{
  \Delta_{\rm irr} = n + \epsilon \qquad\qquad
  \Delta_{\rm rel} = s - {6 \over N+8} \epsilon \,.
 }
For higher order expansions in $\epsilon$, we refer the reader to \cite{Missarov:2006in,Missarov:2006iu}. We may naturally suppose that $\Delta_{\rm irr}$ controls the approach of a discrete flow from the free $O(N)$ model to the $p$-adic Wilson-Fisher fixed point, while $\Delta_{\rm rel}$ is the dimension of a mass-like operator which generically drives trajectories away from the fixed point.

\section{Large $N$ methods}
\label{VASILIEV}

Methods based on the Hubbard-Stratonovich transformation have been developed, notably in \cite{Vasiliev:1981yc,Vasiliev:1981dg}, which resum an infinite set of diagrams of the $O(N)$ model at fixed order in large $N$ and allow a determination of critical exponents at the Wilson-Fisher fixed point which are known exactly as functions of $\epsilon$ and to a few orders in large $N$.  Whereas Wilsonian methods are significantly different for field theories defined over $\mathbb{Q}_{p^n}$ than for field theories defined over $\mathbb{R}^n$, the large $N$ methods work nearly identically in the two cases.  We will illustrate this by working out the leading non-trivial results for anomalous dimensions in $\phi^4$ theory.

\subsection{Action}

We start with an informal introduction to the methods of \cite{Vasiliev:1981yc,Vasiliev:1981dg}.  We are interested in a conformally invariant theory, and so we will naively turn off the relevant mass deformation while keeping the $\phi^4$ interaction.  The action \eno{ONmodel} becomes
 \eqn{ONagain}{
  S = \int dx \, \left[ {1 \over 2} \phi^i(x) D^s \phi^i(x) + {\lambda \over 4!} 
    \left( \phi^i(x) \phi^i(x) \right)^2 \right] \,.
 }
Here and below, integrals are over all of $\mathbb{Q}_{p^n}$, or all of $\mathbb{R}^n$, unless indicated otherwise, and $\phi^i$ takes values in $\mathbb{R}^N$.

Acting with $D^s$ in position space is, by definition, the same as multiplying by $|k|^s$ in momentum space:
 \eqn{DsFourier}{
  \int dx \, \chi(kx)^* D^s \phi^i(x) \equiv |k|^s \phi^i(k) \,.
 }
A Fourier integral of fundamental importance is
 \eqn{DsCharacter}{
  \int dk \, \chi(kx) |k|^s = {1/\Gamma(-s) \over |x|^{n+s}} + \hbox{contact terms} \,.
 }
We have previously defined Fourier transforms over $\mathbb{Q}_{p^n}$ in \eno{FourierDef}.  For $\mathbb{R}^n$, we set $\chi(kx) = e^{2\pi i \vec{k} \cdot \vec{x}}$.  Thus, relative to standard conventions in quantum field theory, our wave numbers $\vec{k}$ always include an extra factor of $1/2\pi$.\footnote{Restoring dimensions by writing a plane wave as $e^{i\vec{p} \cdot \vec{x} / \hbar}$, where now $\vec{p}$ is the momentum, the current conventions can be understood as arising from setting $h\equiv 2\pi\hbar = 1$ rather than following the usual practice of setting $\hbar = 1$.\label{hbar}}

In the case of $p$-adic numbers, a sufficient prescription for the contact terms is for them to be just a delta function, so that we recover the Vladimirov derivative:
 \eqnQq{VladimirovD}{
  D^s \phi^i(x) \equiv {1 \over \Gamma(-s)}
    \int dy \, {\phi^i(y) - \phi^i(x) \over |y-x|^{n+s}} \,.
 }
Some of the good properties of the Vladimirov derivative are explained, for instance, in Appendix B of \cite{Gubser:2016htz}.  The Vladimirov derivative should be understood to act on functions which can be approximated as piecewise constant functions with compact support.

In the case of $\mathbb{R}^n$, the contact terms have in general a more complicated structure, including both delta functions and derivatives of delta functions.  At a formal level, we can let $s$ remain a continuously variable parameter in the real case.  The theories so obtained have bilocal terms in position space, as in \cite{Fisher:1972zz,Sak:1973a}.  When $s$ is a positive even integer, we recover locality:
 \eqnRn{DtwoLap}{
  D^s \phi^i(x) = \square^{s/2}\phi^i(x) \qquad\hbox{for $s=2,4,6,\ldots$}\,,
 }
where 
 \eqnRn{LapDef}{
  \square \equiv -{1 \over (2\pi)^2} \sum_{j=1}^n \partial_j^2 \,.
 }
The general expression \eno{DsFourier} is consistent with \eno{DtwoLap} because $1/\Gamma_{\mathbb{R}^n}(-s)$ has zeros at $s=2,4,6,\ldots$\,.  (Actually, \eno{DtwoLap} is equally valid at $s=0$, where $1/\Gamma_{\mathbb{R}^n}(-s)$ also has a zero, but this is not an interesting case because then the ``kinetic'' term is identical to the mass term.)\footnote{The explicit factors of $2\pi$ in \eno{LapDef} imply a normalization of the kinetic term that is different from the one normally used in field theory: For $\mathbb{R}^n$ with $s=2$, our kinetic term is $S_{\rm kin} = \int_{\mathbb{R}^n} dx \, {1 \over 8\pi^2} (\partial\phi^i)^2$ instead of the more standard $S_{\rm kin} = \int_{\mathbb{R}^n} dx \, {1 \over 2} (\partial\phi^i)^2$.  This means that our field $\phi^i$ includes an extra factor of $2\pi$ compared with standard conventions, and as a result, powers of $2\pi$ will show up in all our position space Green's functions that do not match the literature.  More precisely: explicit factors of $2\pi$ altogether disappear from Green's functions when we follow our conventions faithfully, including the use of $\Gamma_{\mathbb{R}^n}$ (defined in \eno{GammaBeta} and \eno{zetaDefRn}) rather than $\Gamma_{\rm Euler}$.\label{FieldNormalization}}

The Hubbard-Stratonovich trick is to replace
 \eqn{HStrick}{
  {\lambda \over 4!} (\phi^i \phi^i)^2 \to 
    {\lambda \over 4!} (\phi^i \phi^i)^2 - 
       {3 \over 2\lambda N} \left( \sigma - {\lambda\sqrt{N} \over 6} \phi^i \phi^i \right)^2
   = {1 \over 2 \sqrt{N}} \sigma\phi^i \phi^i - {3\sigma^2 \over 2\lambda N}
 }
in the action.  This is permitted because we can eliminate $\sigma$ by its equation of motion and recover the original action.  At the level of path integration the same manipulation is still permitted, but $\sigma$ must run over imaginary rather than real values in order to have a convergent integral in the $\sigma$ direction.  Next we assume that $\lambda N$ runs to large values, so that the $\sigma^2 / \lambda N$ term in \eno{HStrick} may be neglected.  Thus we arrive at the modified action
 \eqn{Smod}{
  S = \int dx \, \left[ {1 \over 2} \phi^i(x) D^s \phi^i(x) + 
    {1 \over 2\sqrt{N}} \sigma\phi^i \phi^i \right] \,.
 }
We may alternatively understand \eno{Smod} as arising from a non-linear sigma model where for each $x$, $\phi^i(x)$ is constrained to lie on a sphere $S^{N-1}$ of fixed radius; then $\sigma$ is the Lagrange multiplier that enforces the constraint, and there is an extra term linear in $\sigma$ whose role is to fix the radius of the $S^{N-1}$---or in diagrammatic terms, to eliminate any tadpole for $\sigma$.

\subsection{Leading order propagators}
\label{LEADING}

\begin{figure}[h!]
		\center
			\includegraphics{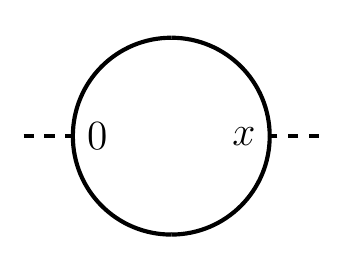}
			\caption{The vacuum polarization diagram for the $\sigma$ field. Dashed lines correspond to the $\sigma$ field, and solid lines correspond to the $\phi$ field.}			\label{VacuumPolarization}
		\end{figure}

A two-point function for $\phi^i$ can be read off from \eno{Smod} at tree level:
 \eqn{GammaPhi}{
  \Gamma^{(0)}_{\phi\phi}(k) = |k|^s \qquad\qquad
  G^{(0)}_{\phi\phi}(k) = {1 \over |k|^s} \qquad\qquad
  G^{(0)}_{\phi\phi}(x) = {1 /\Gamma(s) \over |x|^{n-s}} \,.
 }
All these two-point functions include a factor of $\delta^{ij}$ which we suppress.  All position space correlators should be understood as subject to correction by contact terms.  The 1PI two-point amplitude for $\sigma$ gets its first contribution at one loop as shown in figure~\ref{VacuumPolarization}:
 \eqn{GammaSigma}{
  \Gamma^{(1)}_{\sigma\sigma}(x) = -{1/\Gamma(s)^2 \over 2 |x|^{2n-2s}} \,.
 }
The explicit sign in \eno{GammaSigma} comes from the convention that field configurations are weighted by $e^{-\Gamma}$.  The $1/2$ is a symmetry factor, and the rest of the amplitude is the square of $G^{(0)}_{\phi\phi}(x)$.  A factor of $N$ for the sum over indices in the $\phi^i$ loop is offset by two factors of $1/\sqrt{N}$, one from each vertex.  Straightforward Fourier transforms lead to
 \eqn{GSigma}{
  G^{(1)}_{\sigma\sigma}(k) = -{2 \over {\rm B}(n-s,n-s)} |k|^{2s-n} \qquad\qquad
  G^{(1)}_{\sigma\sigma}(x) = -2 {\Gamma(2s) \over {\rm B}(n-s,n-s)} 
   {1 \over |x|^{2s}} \,.
 }

From $G^{(0)}_{\phi\phi}(x) \propto 1/|x|^{n-s}$ we conclude $\Delta_\phi = {n-s \over 2} + {\cal O}(1/N)$, which for $\mathbb{Q}_{p^n}$ trivially agrees with the conclusion of the non-renormalization theorem of section~\ref{NONRENORMALIZATION}, which indicates that $\Delta_\phi$ receives no corrections from its free field value.  This agreement is trivial because we're only looking at tree-level contributions to $G_{\phi\phi}(x)$ thus far.

From $G^{(0)}_{\sigma\sigma}(x) \propto 1/|x|^{2s}$ we conclude $\Delta_\sigma = s + {\cal O}(1/N)$.  We identify $\sigma$ itself as the relevant deformation, so from \eno{TwoDimensions} we see that we already have agreement between $\Delta_{\rm rel}$ and $\Delta_\sigma$ to leading order in small $\epsilon$ and large $N$.  Our computations in section~\ref{DELTASIGMA} will extend this agreement to the next order: that is, we will find
 \eqn{DeltaSigma}{
  \Delta_\sigma = s - {6 \over N} \epsilon + {\cal O}(1/N^2) + {\cal O}(\epsilon^2) \,.
 }
First, however, we will show that on the $p$-adics $\Delta_\phi$ receives no correction through ${\cal O}(1/N)$.

\subsection{Self-energy diagram I: Momentum space methods}
\label{SELFONE}

\begin{figure}[h!]
		\center
			\includegraphics{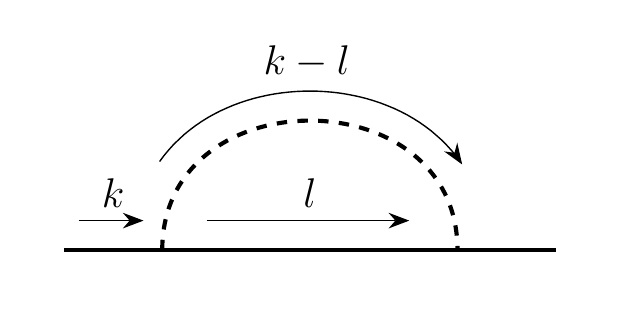}
			\caption{The self energy diagram for the $\phi$ field.}			\label{SelfEnergy}
		\end{figure}

The self-energy correction to the 1PI two-point function for $\phi^i$ is given by the diagram in figure~\ref{SelfEnergy}, whose amplitude is
 \eqn{GTwoK}{
  \Gamma^{(2)}_{\phi\phi}(k) = 
    -{1 \over N} \int d\ell \, G^{(0)}_{\phi\phi}(\ell) G^{(1)}_{\sigma\sigma}(k-\ell)
     = {2/N \over {\rm B}(n-s,n-s)}
        \int d\ell \, |\ell|^{-s} |k-\ell|^{2s-n} \,.
 }
 
Evidently we must investigate the divergence properties of the integral.  It helps to introduce functions
 \eqn{Multiplicative}{
  \pi_t(k) \equiv |k|^{t-n}
 }
for any complex number $t$.  These functions are multiplicative characters on $\mathbb{Q}_{p^n}$,\footnote{Given a field $K$, a multiplicative character $\pi: K^\times \to \mathbb{C}^\times$ satisfies $\pi(xy) = \pi(x) \pi(y)$.} but are also of course well defined on $\mathbb{R}^n$ despite there being no obvious notion of multiplicative characters there (unless $n=1$ or $2$, where we have $\mathbb{R}$ or $\mathbb{C}$ respectively, both of which are fields).  The Fourier transform of $\pi_t(k)$ is $\hat\pi_t(x) \equiv \Gamma(t) |x|^{-t}$ up to contact terms, so the obvious identity $\hat\pi_{t_1}(x) \hat\pi_{t_2}(x) = {\rm B}(t_1,t_2) \hat\pi_{t_1+t_2}(x)$ becomes in Fourier space
 \eqn{FourierBeta}{
  (\pi_{t_1} * \pi_{t_2})(k) \equiv \int d\ell \, |\ell|^{t_1-n} |k-\ell|^{t_2-n} = 
   {\rm B}(t_1,t_2) \pi_{t_1+t_2}(k) = {\rm B}(t_1,t_2) |k|^{t_1+t_2-n} \,,
 }
The integral in \eno{FourierBeta} converges provided $t_1>0$, $t_2>0$, and $t_1+t_2<n$.  Outside this triangular region, we need to consider some regularization.

Suppose $t_1>0$, $t_2>0$, but $t_1+t_2 > n$, so that the integral in \eno{FourierBeta} has an ultraviolet divergence.  In $\mathbb{Q}_{p^n}$, imposing a hard momentum cutoff leads to
 \eqnQq{CutoffInt}{
  \int_{|\ell| \leq \Lambda} d\ell \, |\ell|^{t_1-n} |k-\ell|^{t_2-n} = 
    {\rm B}(t_1,t_2) |k|^{t_1+t_2-n} + {\zeta(t_1+t_2-n) \over \zeta(n)}
      \Lambda^{t_1+t_2-n} \,,
 }
provided $|k| < \Lambda$.  To obtain \eno{CutoffInt}, the simplest method is to split the integral into regions where $|\ell|$ and $|k-\ell|$ are constant, and then the integral becomes a discrete sum which can be performed exactly.  What is notable about \eno{CutoffInt} is that the result is the sum of two terms: the expression $|k|^{t_1+t_2-n} {\rm B}(t_1,t_2)$ that we got through formal manipulations in \eno{FourierBeta}, plus the $k=0$ result.  Applying \eno{CutoffInt} to \eno{GTwoK}, we now find for $\mathbb{Q}_{p^n}$ the result
 \eqnQq{GTwoReg}{
  \Gamma^{(2)}_{\phi\phi}(k) = {2/N \over {\rm B}(n-s,n-s)}
   \left[ |k|^s {\rm B}(n-s,2s) + {\zeta(s) \over \zeta(n)} \Lambda^s \right] \,.
 }
The divergent piece can be canceled by a counterterm $S_{\rm ct} \propto \int dx \, \Lambda^s \phi^i \phi^i$.  The absence of wave-function renormalization is due to the fact that the divergent part of $\Gamma^{(2)}_{\phi\phi}(k)$ has no $k$-dependence.  In particular, we don't see an anomalous dimension for $\phi^i$ (at this level) because there is no term proportional to $|k|^s \log(\Lambda/|k|)$.  There is only a finite renormalization of the two-point function for $\phi$:
 \eqnQq{GammaCorrected}{
  \Gamma^{(0)}_{\phi\phi}(k) + \Gamma^{(2)}_{\phi\phi}(k) = 
   \left[ 1 + {2 \over N} {{\rm B}(n-s,2s) \over {\rm B}(n-s,n-s)} \right] |k|^s \,.
 }

The case of $\mathbb{R}^n$ is harder because there is no such exact formula as \eno{CutoffInt}, owing to the possibility of subleading divergences.  Focusing on the case where the leading divergence is quadratic,
 \eqnRn{CutoffReal}{
  \int_{|\ell| \leq \Lambda} d\ell \, |\ell|^{t_1-n} |k-\ell|^{2-t_1} &= 
    {2 \over \zeta(n)} \left[ {\Lambda^2 \over 2} - {(t_1-n)(2-t_1) \over 2n}
      k^2 \log {\Lambda \over |k|} + \hbox{(finite)} \right]  \cr
   &= {(t_1-n)(2-t_1) \over n\zeta(n)} k^2 \log |k| + \hbox{(non-universal)} \,,
 }
where $k^2 = |k|^2 = \sum_{i=1}^n k_i^2$ and we restrict $0<t_1<n+2$ to avoid infrared divergences.  In \eno{CutoffReal}, ``finite'' means terms which remain finite as $\Lambda \to \infty$ with $k$ held fixed.  The precise way in which we impose the cutoff doesn't affect the terms shown; for instance, we could have integrated instead over the region $|k-\ell| \leq \Lambda$.  The logarithmic term is particularly robust, in that even a rescaling of $\Lambda$ does not affect it.  This is the familiar scheme independence of leading logarithmic terms, which we emphasize in the second line by picking out the $k^2 \log |k|$ behavior explicitly and folding the $k^2 \log \Lambda$ term along with the $\Lambda^2$ term into the ``non-universal'' part.  These divergent terms can be canceled by local counterterms.  Of course, all this is textbook renormalization procedure, worthy of note here only as a segue into a more formal method to extract the same leading logarithmic term which will generalize conveniently to the $p$-adic context in section~\ref{DELTASIGMA}.  This more formal method is to ``regularize'' by shifting one of the exponents of the integral and then treating that shift as small:
 \eqnRn{RegBeta}{
  \int d\ell \, |\ell|^{t_1-n} |k-\ell|^{2-t_1-\delta} &= 
   {\rm B}(t_1,n+2-t_1-\delta) |k|^{2-\delta} = 
   {\rm B}(t_1,-2+\delta) |k|^{2-\delta} \cr
   &= {(t_1-n)(2-t_1) \over n \zeta(n)} \left[ -{1 \over \delta} + \log |k| \right] k^2 + 
   \hbox{(finite)} \,.
 }
The first line of \eno{RegBeta} is rigorously valid when $t_1-n-2 < \delta < -2$.  To reach the second line of \eno{RegBeta}, we analytically continue in $\delta$ past the singularity of ${\rm B}(t_1,-2+\delta)$ at $\delta = -2$ to the next singularity, at $\delta=0$.  Evidently, the $k^2 \log |k|$ term matches what was found in \eno{CutoffReal}.  In applications of \eno{RegBeta} and related analytic continuations to diagrammatic amplitudes, we must be careful to shift dimensions at the level of the Feynman rules.  Our choice is to shift exponents associated with the $\sigma$ propagator.

The contrast between $\mathbb{R}^n$ and $\mathbb{Q}_{p^n}$ is clear from the position of poles in the beta function.  If we tried the same manipulation as \eno{RegBeta} for the $p$-adics, we would get a finite result and no $\log |k|$ term because there is no singularity in ${\rm B}_{\mathbb{Q}_{p^n}}(t_1,-2+\delta)$ at $\delta=0$; but ${\rm B}_{\mathbb{R}^n}(t_1,-2+\delta)$ does have such a pole on account of the infinite sequence of poles in $\Gamma_{\mathbb{R}^n}(t)$.\footnote{It is intriguing to note that the same contrast between analytic properties of $\Gamma_{\mathbb{R}}$ and $\Gamma_{\mathbb{Q}_p}$ is responsible for the presence of infinitely many states in the Archimedean string spectrum, whereas the standard $p$-adic string construction gives only a tachyon.  We will return to this line of thought further in section~\ref{DISCUSSION}.}  If, on the other hand, we were considering an integral like $\int d\ell \, |\ell|^{t_1-n} |k-\ell|^{-t_1}$ which is logarithmically divergent, then a $\log |k|$ term would come out of any sensible regularization procedure regardless of whether the integral is over $\mathbb{R}^n$ or $\mathbb{Q}_{p^n}$.  In the approach where we shift one exponent, the $\log |k|$ term would be associated with a pole in ${\rm B}(t_1,\delta)$ at $\delta=0$, which is present equally for ${\rm B}_{\mathbb{R}^n}$ and ${\rm B}_{\mathbb{Q}_{p^n}}$.

With \eno{CutoffReal} or \eno{RegBeta} in hand, we can calculate the anomalous dimension for $\phi^i$ in the standard setup of a local field theory on $\mathbb{R}^n$: Setting $s=2$ and keeping only the universal leading logarithmic term, we have
 \eqnRn{GammaToAnomalous}{
  \Gamma^{(0)}_{\phi\phi}(k) + \Gamma^{(2)}_{\phi\phi}(k) = 
    k^2 - {4/N \over {\rm B}(n-2,n-2)} {4-n \over n\zeta(n)} 
     k^2 \log |k| = |k|^{n-2\Delta_\phi} \,,
 }
where
 \eqnRn{AnomalousPhi}{
  \Delta_\phi = {n-2 \over 2} + {2/N \over {\rm B}(n-2,n-2)} 
      {4-n \over n \zeta(n)} + {\cal O}(1/N^2) \,.
 }
This result is exact in $\epsilon = 4-n$, but if we wish to compare with standard perturbation theory we can expand in small $\epsilon$:
 \eqnRn{AnomalousSmall}{
  \Delta_\phi = {n-2 \over 2} + {\epsilon^2 \over 4 N} + {\cal O}(\epsilon^3) + 
    {\cal O}(1/N^2) \,.
 }

It is possible to unify our perspective somewhat by writing a formula for $\Delta_\phi$ which is valid equally for $\mathbb{R}^n$ and $\mathbb{Q}_{p^n}$:
 \eqn{DoublePurpose}{
  \Delta_\phi = {n-s \over 2} + {1 \over N}
   \Res_\delta {{\rm B}(n-s,-s+\delta) \over {\rm B}(n-s,n-s)} + {\cal O}(1/N^2) \,,
 }
where we understand $\Res_z$ as picking out the residue at a pole at $z=0$ of a meromorphic function of $z$:
 \eqn{ResDef}{
  \Res_z f(z) \equiv \oint_0 {dz \over 2\pi i} \, f(z) \,.\
 }
This unified perspective suggests in $\mathbb{R}^n$ that $s=2$ may not be as special as we normally think---and that in particular, any positive even $s$ will give rise to constructions similar to the Wilson-Fisher fixed point, obtained (one might assume) from local Gaussian theories by adding a $\phi^4$ term.  We follow up this idea in section~\ref{HIGHER}.  When applied to $\mathbb{Q}_{p^n}$ (assuming $s>0$), \eno{ResDef} tells us correctly that the anomalous dimension vanishes since ${\rm B}(n-s,-s+\delta)$ is finite at $\delta=0$.  Although the expression \eno{DoublePurpose} appears to be merely a repackaging of previous results, it does highlight the origin of the anomalous dimension and suggests the possibility of extending to more general base fields and/or more interesting multiplicative characters.

\subsection{Self-energy diagram II: Position space methods}
\label{SELFTWO}

The evaluation of the 1PI self-energy diagram is trivial in position space:
 \eqn{GammaTwoEval}{
  \Gamma^{(2)}_{\phi\phi}(x) = -{1 \over N} G^{(0)}_{\phi\phi}(x) G^{(1)}_{\sigma\sigma}(x) = 
    {2 \over N} {\Gamma(2s)/\Gamma(s) \over {\rm B}(n-s,n-s)} {1 \over |x|^{n+s}} \,.
 }
In $\mathbb{Q}_{p^n}$ we can straightforwardly combine $\Gamma^{(2)}_{\phi\phi}(x)$ with $\Gamma^{(0)}_{\phi\phi}(x) = {1/\Gamma(-s) \over |x|^{n+s}}$ to obtain the finite renormalization factor appearing already in \eno{GammaCorrected}.  In $\mathbb{R}^n$ this fails because $\Gamma^{(0)}_{\phi\phi}(x) = \square \delta(x)$.  A more effective method is to investigate the contribution of the self-energy graph to the connected two-point function:
 \eqn{GTwoEval}{
  G^{(2)}_{\phi\phi}(x) = -{2/N \over \Gamma(s) {\rm B}(s,s) {\rm B}(n-s,n-s)} 
    I_3(x) \,,
 }
where we define
 \eqn{Ithree}{
  I_3(x) \equiv \int dx_1 dx_2 \, {1 \over |x_1|^{n-s} |x_{12}|^{n+s-\delta} |x-x_2|^{n-s}}
   \,,
 }
where $x_{12} = x_1-x_2$.  Anticipating possible divergences, we've already introduced as a regulator a shift $\delta$ in one of the exponents.  We have coordinated the normalization of $\delta$ in \eno{Ithree} with the normalization we used in \eno{RegBeta}: in both cases, we're effectively sending $\Delta_\sigma \to \Delta_\sigma - \delta/2$ while holding all other quantities fixed.

Because $I_3(x)$ is the convolution of three power laws, it is easily evaluated using \eno{FourierBeta}.  (We don't mean to pass to Fourier space; we mean to apply \eno{FourierBeta} as is with $k$ variables replaced with $x$ variables.)  The result is
 \eqn{IThreeValue}{
  I_3(x) = {{\rm B}(s,s) {\rm B}(2s,-s+\delta) \over |x|^{n-s-\delta}} 
   \deltaeq {{\rm B}(s,s) {\rm B}(n-s,-s+\delta) \over |x|^{n-s-\delta}}.
 }
In the second step, $\deltaeq$ means that the last expression differs from the first only by terms which are finite as $\delta \to 0$.  In the current case, this delta-equality is true provided $s$ avoids special values such as $0$, $n$, and $n/2$.  Thus we arrive at
 \eqn{GTwoValue}{
  G^{(0)}_{\phi\phi}(x) + G^{(2)}_{\phi\phi}(x) &\deltaeq
   {1 / \Gamma(s) \over |x|^{n-s}} \left[ 1 - {2 \over N} {{\rm B}(n-s,-s+\delta) \over
     {\rm B}(n-s,n-s)} |x|^\delta \right]
      \cr
   &\deltaeq {1 / \Gamma(s) \over |x|^{n-s}} \left[ 1 - 
     {2 \over N} 
     \left( \Res_\delta {{\rm B}(n-s,-s+\delta) \over {\rm B}(n-s,n-s)} \right) 
     \left( {1 \over \delta} + \log |x| \right) 
      \right] \,.
 }
As before, we drop the divergent $1/\delta$ piece, understanding that its effects can be offset by a local counterterm.  Comparing \eno{GTwoValue} with the expected power law $G_{\phi\phi}(x) \propto 1/|x|^{2\Delta_\phi}$, we arrive at
 \eqn{DeltaPhiAgain}{
  \gamma_\phi \equiv \Delta_\phi - {n-s \over 2} 
    = {1 \over N} \Res_\delta {{\rm B}(n-s,-s+\delta) \over {\rm B}(n-s,n-s)} + 
     {\cal O}(1/N^2) \,.
 }
This is easily seen to agree with \eno{DoublePurpose} provided we stipulate $s>0$.  For $\mathbb{R}^n$ (and $s=2$ as we always stipulate for the Archimedean case) it also agrees with the standard result \cite{Vasiliev:1981yc,Vasiliev:1981dg}
 \eqnRn{DeltaPhiRn}{
  \gamma_\phi = {n-4 \over N} {2^{n-3} \over \pi^{3/2}} {
   \Gamma_{\rm Euler}\left( {n-1 \over 2} \right) \over
    \Gamma_{\rm Euler}\left( {n \over 2} + 1 \right)} \sin {\pi n \over 2} + {\cal O}(1/N^2) \,.
 }

\subsection{Corrections to the $\sigma$ propagator}
\label{DELTASIGMA}

\begin{figure}[h!]
		\center
			\includegraphics{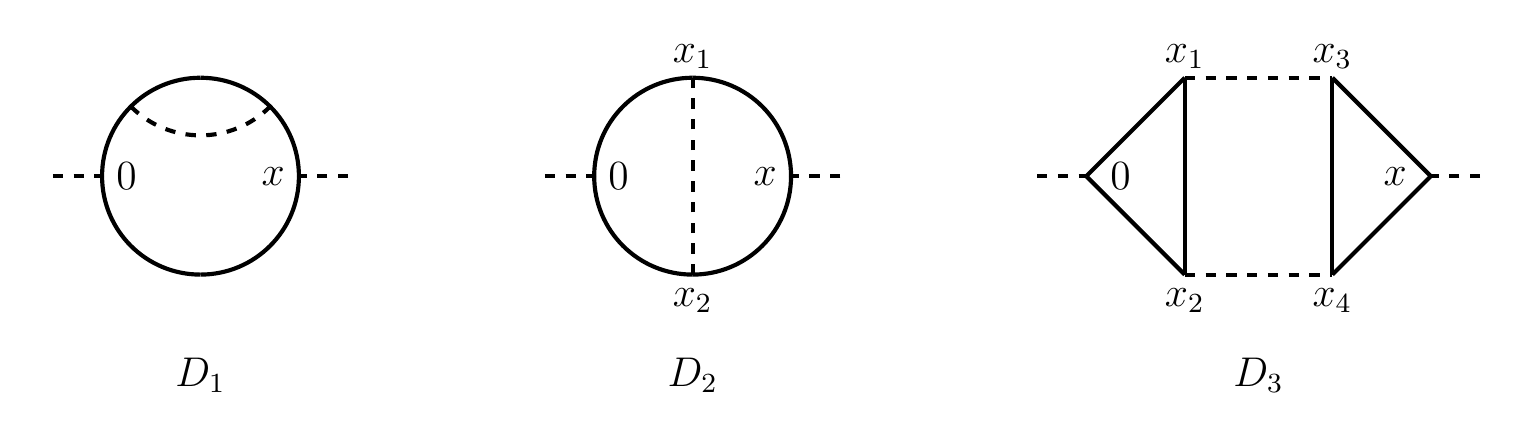}
			\caption{The three position space diagrams that contribute to $1/N$ corrections to the anomalous dimension of the $\sigma$ field.}			\label{ThreeDiagrams}
		\end{figure}

In order to arrive at \eno{DeltaSigma}, we need to find contributions to $\Gamma_{\sigma\sigma}(x)$ at order $1/N$.\footnote{After the discussion of section~\ref{SELFTWO} one might expect that carrying through to $G_{\sigma\sigma}(x)$ is necessary in order to avoid comparing power laws to contact terms in the case of $\mathbb{R}^n$.  This is not a problem because in $\Gamma^{(1)}_{\sigma\sigma,\mathbb{R}^n}(x) \propto 1/|x|^{2n-4}$ we allow ourselves to analytically continue in $n$---and the only points of concern are the upper and lower critical dimensions, $n=4$ and $2$.}  There are three diagrams which contribute: $D_1$, $D_2$, and $D_3$ as shown in figure~\ref{ThreeDiagrams}.  The first is easy because the only logarithmic divergence arises from the self-energy subdiagram, and it can be tracked by replacing the two-loop diagram with the one-loop diagram in figure~\ref{VacuumPolarization}, only with the tree-level propagators $G^{(0)}_{\phi\phi}(x)$ replaced by
 \eqn{GTwoOnceMore}{
  G^{(0)}_{\phi\phi}(x) + G^{(2)}_{\phi\phi}(x) \deltaeq
    {1/\Gamma(s) \over |x|^{n-s}} \left[ 1 - 2 \gamma_\phi 
      \left( {1 \over \delta} + \log |x| \right) \right] \,,
 }
where we have rewritten \eno{GTwoValue} in compact form.  We remember that $\gamma_\phi$ is ${\cal O}(1/N)$ and vanishes for $\mathbb{Q}_{p^n}$.  Thus, following through the manipulations of section~\ref{LEADING}, we find
 \eqn{GammaWithDone}{
  \Gamma^{(1)}_{\sigma\sigma}(x) + \Gamma^{(D_1)}_{\sigma\sigma}(x) \deltaeq 
   -{1/\Gamma(s)^2 \over 2|x|^{2n-2s}}
     \left[ 1 - 4 \gamma_\phi \left( {1 \over \delta} + \log |x| \right) \right] \,,
 }
implying that diagram $D_1$ contributes $\gamma^{(D_1)}_\sigma = -2\gamma_\phi$ to the anomalous dimension
 \eqn{gammaSigma}{
  \gamma_\sigma \equiv \Delta_\sigma - s \,.
 }
To get the contributions to $\gamma_\sigma$ from $D_2$ and $D_3$ we need only isolate their leading logarithmic terms and add those terms to \eno{GammaWithDone}.

The second diagram contributes
 \eqn{GammaDtwo}{
  \Gamma^{(D_2)}_{\sigma\sigma}(x) = -{1 \over 2N} \left( {1 \over \Gamma(s)} \right)^4 
    \left( -2 {\Gamma(2s) \over {\rm B}(n-s,n-s)} \right) I_{D_2}(x) \,.
 }
The leading sign is the usual one for 1PI diagrams; the $1/2$ is a symmetry factor; $1/N$ comes from index summation together with four $\sigma\phi\phi$ vertices; the remaining prefactors come from the four $G^{(0)}_{\phi\phi}$ propagators and the one internal $G^{(1)}_{\sigma\sigma}$ propagator; and
 \eqn{IDtwo}{
  I_{D_2}(x) &= \int dx_1 dx_2 \, {1 \over |x_1|^{n-s} |x_1-x|^{n-s} |x_{12}|^{2s-\delta}
     |x_2|^{n-s} |x_2-x|^{n-s}}  \cr
   &\deltaeq {\rm B}(s,s) {\rm B}(\delta,\delta) |x|^{2s-2n+\delta} \,.
 }
The second equality in \eno{IDtwo} takes a little work to justify, and we postpone a derivation to section~\ref{SHIFT}.  Combining \eno{GammaDtwo} and \eno{IDtwo} we see that
 \eqn{GammaDtwoValue}{
  \Gamma^{(D_2)}_{\sigma\sigma}(x) &\deltaeq {1 \over N} 
    {{\rm B}(\delta,\delta) \over \Gamma(s)^2 {\rm B}(n-s,n-s)} |x|^{2s-2n+\delta}  \cr
   &\deltaeq {1/N \over \Gamma(s)^2} |x|^{2s-2n} 
    \left( \Res_\delta {{\rm B}(\delta,\delta) \over {\rm B}(n-s,n-s)} \right) 
     \left[ {1 \over \delta} + \log |x| \right] \,,
 }
from which we deduce in turn the contribution to the anomalous dimension
 \eqn{AnomalousTwo}{
  \gamma^{(D_2)}_\sigma &= -{1 \over N}   
    \Res_\delta {{\rm B}(\delta,\delta) \over {\rm B}(n-s,n-s)} \,.
 }

The third diagram contributes
 \eqn{GammaDthree}{
  \Gamma^{(D_3)}_{\sigma\sigma}(x) = -{1 \over 2N}
   \left( {1 \over \Gamma(s)} \right)^6 
   \left( -2 {\Gamma(2s) \over {\rm B}(n-s,n-s)} \right)^2 I_{D_3}(x)
 }
where
 \eqn{IDthree}{
  I_{D_3}(x) &= \int dx_1 dx_2 dx_3 dx_4 \, {1 \over |x_1|^{n-s} |x_2|^{n-s}
    |x_{12}|^{n-s}} \times {1 \over |x_{13}|^{2s-\delta/2} |x_{24}|^{2s-\delta/2}}  \cr
   &\qquad\qquad{} \times {1 \over |x_3-x|^{n-s} |x_4-x|^{n-s} |x_{34}|^{n-s}}  \cr
   &\deltaeq {\rm B}(s,s)^2 {\rm B}(n-s,n-2s) {\rm B}(\delta,\delta) |x|^{2s-2n+\delta} \,.
 }
The first and third factors in the integrand of \eno{IDthree} come from the $G^{(0)}_{\phi\phi}$ propagators running around the triangular loops.  The second factor comes from the internal $G^{(1)}_{\sigma\sigma}$ propagators.\footnote{The alert reader may be surprised that we chose $\Delta_\sigma \to \Delta_\sigma - \delta/4$ as a regulator in the $G^{(1)}_{\sigma\sigma}$ propagators in \eno{IDthree}, in contrast to our previous strategy $\Delta_\sigma \to \Delta_\sigma - \delta/2$.  We made this new choice because there are two $G^{(1)}_{\sigma\sigma}$ propagators, and we wanted the added $x$ dependence arising from the regulator to be $|x|^\delta$ rather than $|x|^{2\delta}$.  Our new choice does not affect the leading logarithmic term: The leading terms in a small $\delta$ expansion involve a factor ${1 \over \delta} + \log |x|$, whereas if we had stuck with $\Delta_\sigma \to \Delta_\sigma - \delta/2$ we would have found ${1 \over 2\delta} + \log |x|$.}   \eqn{GammaDthreeValue}{
  \Gamma^{(D_3)}_{\sigma\sigma}(x) &\deltaeq -{2/N \over \Gamma(s)^2} 
    {{\rm B}(n-s,n-2s) {\rm B}(\delta,\delta) \over {\rm B}(n-s,n-s)^2}
    |x|^{2s-2n+\delta}  \cr
   &\deltaeq -{2/N \over \Gamma(s)^2} |x|^{2s-2n}
    \left( \Res_\delta {{\rm B}(n-s,n-2s) {\rm B}(\delta,\delta) \over {\rm B}(n-s,n-s)^2} \right) 
     \left[ {1 \over \delta} + \log |x| \right] \,,
 }
from which we deduce in turn
 \eqn{AnomalousThree}{
  \gamma^{(D_3)}_\sigma &= {2 \over N} \Res_\delta 
   {{\rm B}(n-s,n-2s) {\rm B}(\delta,\delta)\over {\rm B}(n-s,n-s)^2} \,.
 }
Putting the contributions from $D_1$, $D_2$, and $D_3$ together, we arrive at the anomalous dimension
 \eqn{gammaSigmaRes}{
  \gamma_\sigma &= \gamma^{(D_1)}_\sigma + \gamma^{(D_2)}_\sigma + \gamma^{(D_3)}_\sigma + 
    {\cal O}(1/N)^2  \cr
   &= {1 \over N} \Res_\delta \left[ -2 {{\rm B}(n-s,-s+\delta) \over {\rm B}(n-s,n-s)} + 
     \left( -1 + 2 {{\rm B}(n-s,n-2s) \over {\rm B}(n-s,n-s)} \right) 
       {{\rm B}(\delta,\delta) \over {\rm B}(n-s,n-s)} \right]   \cr
   &\qquad{} + {\cal O}(1/N^2) \,.
 }

The first term in square brackets comes from $D_1$ and vanishes for $\mathbb{Q}_{p^n}$. For $\mathbb{R}^n$ and $s=2$ we recover from \eno{gammaSigmaRes}  the result of \cite{Vasiliev:1981yc,Vasiliev:1981dg}:
 \eqnRn{gammaSigmaRn}{
  \gamma_\sigma = 4 {(n-1)(n-2) \over n-4} \gamma_\phi + {\cal O}(1/N^2) \,.
 } 
If we pass to the limit of small $\epsilon$, \eno{gammaSigmaRes} becomes
 \eqn{gammaSigmaLim}{
  \gamma_\sigma = -{6 \over N} \epsilon + {\cal O}(1/N^2) + {\cal O}(\epsilon^2) \,.
 }
 The result \eno{gammaSigmaLim} is valid equally for $\mathbb{R}^n$ and $\mathbb{Q}_{p^n}$, and for $\mathbb{Q}_{p^n}$ we see that it agrees with \eno{DeltaSigma}. If one further expands \eno{gammaSigmaRes} to third order in $\epsilon$ for $\mathbb{Q}_{p^n}$, the result will be found to agree with the $\epsilon$ expansion in \cite{Missarov:2006in,Missarov:2006iu}. If instead we expand about the lower critical dimension and define $\tilde{\epsilon}=n-s$, then equation \eno{gammaSigmaRes} says that
 \eqn{gammaSigmaLower}{
  \gamma_\sigma = \mathcal{O}(\tilde{\epsilon}^2) \,.
 }
This result is also valid equally for $\mathbb{R}^n$ and $\mathbb{Q}_{p^n}$, though the agreement is non-trivial: different terms in \eno{gammaSigmaRes} cancel to make the term linear in $\tilde{\epsilon}$ vanish.

\subsection{Position space integrals I: The star-triangle identity}

Two useful tools for evaluating position space diagrams are the convolution integral \eno{FourierBeta}, which we rewrite here:
 \eqn{FBagain}{
  \int dy \, |x|^{t_1-n} |y-x|^{t_2-n} = {\rm B}(t_1,t_2) |x|^{t_1+t_2-n} \,,
 }
and the star-triangle identity of \cite{Parisi:1971a},\footnote{Originally in \cite{Parisi:1971a} the star-triangle identity was stated for ${\mathbb{R}}^3$ as 
 $$
  \int d^3 t \, |t-x|^a |t-y|^b |t-z|^c = 
   \pi^{3/2} {\Gamma_{\rm Euler}({a+3 \over 2}) \Gamma_{\rm Euler}({b+3 \over 2}) 
     \Gamma_{\rm Euler}({c+3 \over 2}) 
      \over \Gamma_{\rm Euler}(-a/2) \Gamma_{\rm Euler}(-b/2) \Gamma_{\rm Euler}(-c/2)}
       |x-y|^{-3-c} |y-z|^{-3-a} |z-x|^{-3-b}
 $$
provided $a+b+c=-6$.  The somewhat complicated prefactor is precisely ${\rm B}_{\mathbb{R}^3}(a+3,b+3)$.} which can be written compactly as
 \eqn{StarTriangle}{
  \int dx \, \prod_{i=1}^3 |x-x_i|^{t_i-n} = {\rm B}(t_1,t_2) \prod_{i=1}^3 |y_i|^{-t_i}
   \qquad\hbox{if}\qquad \sum_{i=1}^3 t_i = n \,,
 }
where we define
 \eqn{yiDef}{
  y_1 \equiv x_{23} \qquad y_2 \equiv x_{31} \qquad y_3 \equiv x_{12} \,.
 }
The formulas \eno{FBagain}-\eno{StarTriangle} are valid equally for $\mathbb{R}^n$ or $\mathbb{Q}_{p^n}$.  Note that it does not matter which two of $t_1$, $t_2$, and $t_3$ we supply as arguments to ${\rm B}$ in \eno{StarTriangle}.  The integrals \eno{FBagain} and \eno{StarTriangle} are rigorously valid only when the integrals converge.  Provided we set $t_3 = n-t_1-t_2$, the region of convergence for the integrals both in \eno{FBagain} and \eno{StarTriangle} can be characterized by the constraints $t_i > 0$ for all $i$.  Outside this region, we must be prepared to shift exponents (while preserving the constraint $\sum_{i=1}^3 t_i = n$) and cancel divergences against local counterterms, as seen in detail in sections~\ref{SELFONE} and~\ref{SELFTWO} for the self-energy diagram.

In $\mathbb{Q}_{p^n}$, it is possible to evaluate the integral in \eno{StarTriangle} explicitly even when $\sum_{i=1}^3 t_i \neq n$. Due to the ``tall isosceles'' property of ultrametric spaces, for any three non-coincident points $x_1, x_2$ and $x_3$, the linear combinations $y_i$ defined in \eno{yiDef} form the sides of a triangle, such that up to relabeling $y_i$, we always have $|y_1| = |y_2| \geq |y_3|$. With this choice of $y_i$s, the integral in \eno{StarTriangle} can be worked out in general to give
\eqnQq{HypGeoInt}{
  \int  dx  \prod_{i=1}^3 |x-x_i|^{t_i-n} &= {\rm B}(t_1,t_2) |y_2|^{t_3-n} |y_3|^{t_1+t_2-n} \cr  &+ {\rm B}(t_3, t_1+t_2-n) |y_2|^{t_1+t_2+t_3-2n}.
 }
The integral converges provided $t_i >0$ for all $i$, and $t_1 + t_2 + t_3 < 2n$. From the right hand side of \eno{HypGeoInt}, we observe that the integral has poles at $t_i=0$ for all $i$,  at $t_1 + t_2 + t_3 = 2n$, and at  $t_1 + t_2 = n$. Remarkably in $\mathbb{R}^n$, numerics reveal \eno{HypGeoInt} (more precisely the $\mathbb{R}^n$ version constructed from ${\rm B}_{\mathbb{R}^n}$) holds approximately as long as the $L^2$ norms satisfy $|y_1| \approx |y_2| > |y_3|$, although it is no longer an exact identity  like it is in $\mathbb{Q}_{p^n}$.

\subsection{Position space integrals II: Symmetric deformations}
\label{SHIFT}

\begin{figure}[h!]
		\center
			\includegraphics{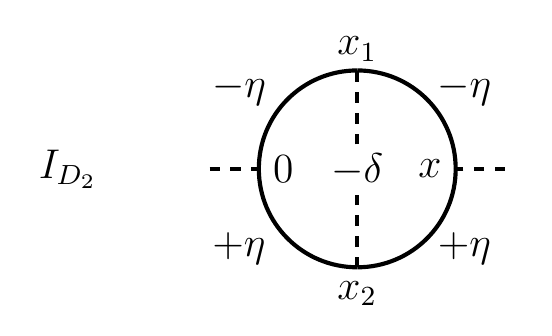}
			\caption{The symmetric deformation of $I_{D_2}$ that allows the integral to be exactly evaluated without disturbing the leading behavior in $\delta$. $\eta$ is eventually set to $\delta/2$.}			\label{SymDeformI2}
		\end{figure}

In order to find the anomalous dimension of the $\sigma$ field by evaluating Feynman diagrams, it is necessary to introduce a regulator to the scaling of the position space $\sigma$ propagator. But when introducing this regulator, the condition $\sum_{i=1}^3 t_i=n$ in equation \eno{StarTriangle} is no longer satisfied, and so the star-triangle identity cannot immediately be applied to equations \eno{IDtwo} and \eno{IDthree}.~\footnote{The more general identity written in \eno{HypGeoInt} can still be employed---we present an alternate derivation of \eno{IDtwo} using this identity in the next section.} There is, however, a way around this obstacle~\cite{Ciuchini:1999wy}. Essentially the idea consists in considering instead of the integrals $I_{D_2}(x)$ and $I_{D_3}(x)$ other integrals that differ from them only by terms that are finite in the $\delta \rightarrow 0$ limit, but to which the star-triangle identity can be applied. Suppose, in \eno{IDtwo}, that we introduce yet another regulator $\eta$ and consider the following integral:
 \eqn{IDtwo2}{
  I_{D_2}(x,\eta) &= \int dx_1 dx_2 \, {1 \over |x_1|^{n-s-\eta} |x_1-x|^{n-s-\eta} |x_{12}|^{2s-\delta}
     |x_2|^{n-s+\eta} |x_2-x|^{n-s+\eta}} .
 }
The deformation is depicted diagrammatically in figure~\ref{SymDeformI2}. Because of the symmetrical manner in which $\eta$ has been introduced, it is clear that $ I_{D_2}(x,\eta)$ is invariant under the transformation $\eta \rightarrow -\eta$. For this reason, and because this Feynman diagram has at most single poles in the regulators, the Taylor expansion of $I_{D_2}(x,\eta)$ in $\eta$ must assume the following form,
 \eqn{IDtwo3}{
  I_{D_2}(x,\eta) &= I_{D_2}(x)+ f_2(x)\eta^2+f_4(x)\eta^4+...
 }
where $f_i(x)$ are some functions that have at most single poles in $\delta$. It is clear then, that if we set $\eta=\frac{\delta}{2}$, then $I_{D_2}(x,\eta)$ will only differ from $I_{D_2}(x)$ by terms that tend to zero as $\delta \rightarrow 0$. But $I_{D_2}(x,\frac{\delta}{2})$ can be evaluated exactly via equations \eno{StarTriangle} and \eno{FBagain}.
 \eqn{IDtwo4}{
  I_{D_2}(x) & \deltaeq  I_{D_2,\delta}(x) =\int  {dx_1 \over |x_1|^{n-s-\frac{\delta}{2}} |x_1-x|^{n-s-\frac{\delta}{2}}}  \, {dx_2 \over  |x_{12}|^{2s-\delta}
     |x_2|^{n-s+\frac{\delta}{2}} |x_2-x|^{n-s+\frac{\delta}{2}}} \cr
     & \deltaeq {\rm B}(s,s)|x|^{2s-n-\delta}\int  {dx_1 \over |x_1|^{n-\delta} |x_1-x|^{n-\delta}} ={\rm B}(s,s){\rm B}(\delta,\delta) |x|^{2s-2n+\delta}.
 } 

\begin{figure}[h!]
		\center
			\includegraphics{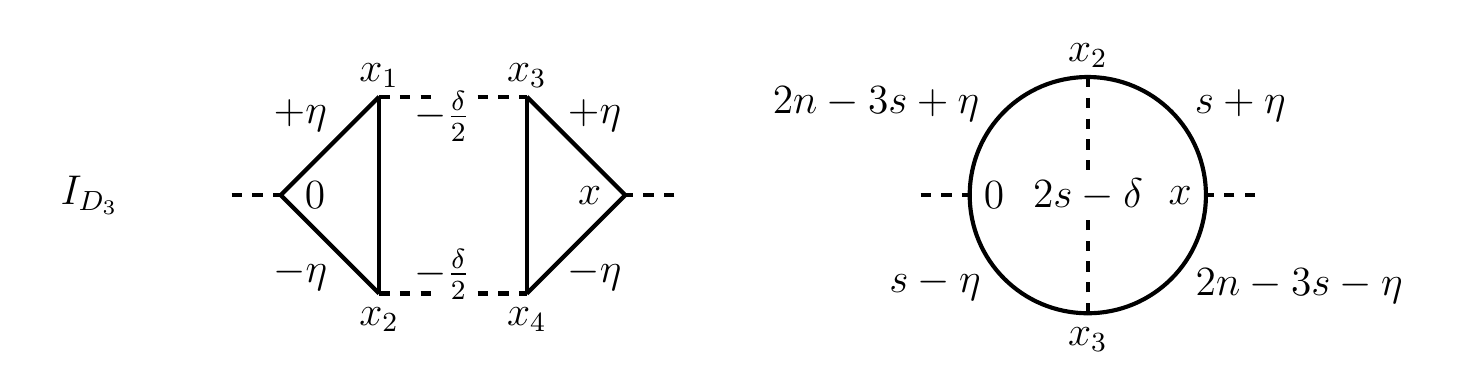}
			\caption{The symmetric deformations that allow $I_{D_3}$ to be exactly evaluated without disturbing the leading behavior in $\delta$. $\eta$ is set to $\delta/2$ in both cases.}			\label{SymDeformI3}
		\end{figure}

This method of finding the leading order behavior of a Feynman diagram by symmetrically changing the scaling of internal propagators and invoking equations \eno{StarTriangle} and \eno{FBagain} can also be used to derive equation \eno{IDthree} in the following manner, represented diagrammatically in figure~\ref{SymDeformI3}:
 \eqn{IDthree2}{
  I_{D_3}(x) & \deltaeq \int {dx_2\, dx_3 \over |x_2|^{n-s-\frac{\delta}{2}}|x_3-x|^{n-s-\frac{\delta}{2}}}  
\int {dx_1\over |x_1|^{n-s+\frac{\delta}{2}}|x_{12}|^{n-s}|x_{13}|^{2s-\frac{\delta}{2}}}  \cr
   &\qquad\qquad{} \times 
    \int {dx_4 \over |x_{24}|^{2s-\frac{\delta}{2}} |x_{34}|^{n-s} |x_4-x|^{n-s+\frac{\delta}{2}}}
   \cr
   & \deltaeq {\rm B}(s,s)^2 \int {dx_2\, dx_3 \over |x_2|^{2n-3s} |x_2-x|^s |x_{23}|^{2s-\delta} |x_3|^s |x_3-x|^{2n-3s}}
   \cr
  & \deltaeq {\rm B}(s,s)^2 \int {dx_3 \over |x_3|^{s-\frac{\delta}{2}}|x_3-x|^{2n-3s-\frac{\delta}{2}}}\, {dx_2 \over |x_2|^{2n-3s+\frac{\delta}{2}}|x_{23}|^{2s-\delta}|x_2-x|^{s+\frac{\delta}{2}}}
   \cr   
&\deltaeq    {\rm B}(s,s)^2{\rm B}(n-s,n-2s)|x|^{2s-n-\delta}\int {dx_3 \over |x_3|^{n-\delta}|x_3-x|^{n-\delta}}
   \cr
   &= {\rm B}(s,s)^2 {\rm B}(n-s,n-2s) {\rm B}(\delta,\delta) |x|^{2s-2n+\delta} \,.
 }
After the second step we recognize the remaining integral as similar to $I_{D_2}$, but with different (and slightly less constrained) exponents.  We represent this diagrammatically on the right side of figure~\ref{SymDeformI3} by showing a diagram with the topology of $D_2$ but with the exponents taken from the second line of \eno{IDthree2}.  The third step, then, is to shift these exponents {\it again} in imitation of how we evaluated $I_{D_2}$.  It may not be entirely evident that the scaling dimensions are altered in a symmetrical manner in the third step in \eno{IDthree2}, but changing variables by letting $x_2 \rightarrow -\tilde{x}_2$ and $\tilde{x}_3 \rightarrow x_3+x$ clearly shows that this is indeed the case.

\subsection{Position space integrals III: Direct evaluation in $\mathbb{Q}_{p^n}$}
We now present an alternate derivation of \eno{IDtwo} which is applicable in $\mathbb{Q}_{p^n}$ and relies on a direct application of the identity in \eno{HypGeoInt}. Using the identity to perform the integral over $x_1$ in \eno{IDtwo}, we obtain
 \eqnQq{IDtwoIntx1}{
 I_{D_2}(x) = \int dx_2\, {1 \over |x_2|^{n-s} |x - x_2|^{n-s}}\, f(|x|,|x_2|,|x-x_2|)\,,
 }
where 
 \eqnQq{fxx2}{
 f(|x|,|x_2|,|x-x_2|) =\begin{cases}   \displaystyle{{{\rm B}(s,s) \over |x_2|^{2s-\delta} |x|^{n-2s}} + {{\rm B}(2s-n,n-\delta) \over |x_2|^{n-\delta}}} & \hbox{ if } |x_2| > |x| \cr\noalign{\vskip2\jot}
 			\displaystyle{{{\rm B}(s,s-\delta) \over |x|^{n-s} |x_2|^{s-\delta}} + {{\rm B}(s,\delta -s)  \over |x|^{n-\delta}}} & \hbox{ if } |x_2| < |x| \cr\noalign{\vskip2\jot}
 			\displaystyle{{{\rm B}(s,s-\delta) \over |x|^{n-s} |x-x_2|^{s-\delta}} + {{\rm B}(s,\delta -s)  \over |x|^{n-\delta}}} & \hbox{ if } |x_2| = |x|\,.
 			\end{cases}
 }
Splitting into the three cases displayed in \eno{fxx2}, the $x_2$ integral in \eno{IDtwoIntx1} is seen to reduce to the following three simple kinds of integrals (with the convergence condition on the exponent shown in parenthesis):
 \eqnQq{SimpleInt}{\seqalign{\span\TL & \span\TR &\qquad\qquad \span\TR}{
 \int_{|y|>|z|} dy\, |y|^{t-n} &= |z|^t {1 \over p^{-t} - 1} \left(1 - {1 \over p^n}\right) & (t<0) \cr
 \int_{|y|<|z|} dy\, |y|^{t-n} &= |z|^t { 1 \over p^t - 1} \left( 1- {1\over p^n}\right) & (t>0)  \cr
 \int_{|y|=|z|} dy\, |y-z|^{t-n} &= |z|^{t} \left(-{1 \over p^n} +  {1 \over 1 - p^{-t}} \left(1 - {1 \over p^n}\right)\right) & (t>0)\,.
 }}
 Plugging \eno{fxx2} in \eno{IDtwoIntx1} and using \eno{SimpleInt} to evaluate the $x_2$ integrals, we end up with the final result
 \eqnQq{IDtwoHypGeoDer}{
 I_{D_2}(x) \deltaeq  {\rm B}(s,s){\rm B}(\delta,\delta) |x|^{2s-2n+\delta}\,,
 } 
 where as usual, $\deltaeq$ means equality up to terms which are finite in the limit $\delta \to 0$.  Though the computation is more cumbersome, the above procedure can also be used to directly evaluate $I_{D_3}(x)$ as well as any other Feynman diagram over the $p$-adics since the integrals always reduce to sums of geometric series.

\section{Higher derivative theories}
\label{HIGHER}

Let us compactly state what has been done so far.  We started with an action
 \eqn{ONOnceMore}{
  S = \int dx \, \left[ {1 \over 2} \phi^i(x) D^s \phi^i(x) + {\lambda \over 4!} 
    \left( \phi^i(x) \phi^i(x) \right)^2 \right]
 }
for Euclidean quantum field theory.  $D^s$ is an $s$-th order derivative operator, implemented by multiplying $\phi$ by $|k|^s$ in Fourier space.  It is assumed that $s>0$.  The sums over $i$ run from $1$ to $N$, which we take to be large.  The integration is over an $n$-dimensional vector space $V$.  We have considered the cases $V = \mathbb{R}^n$ and $V = \mathbb{Q}_{p^n}$, which are $n$-dimensional vector spaces over $\mathbb{R}$ and $\mathbb{Q}_p$, respectively.  $\mathbb{Q}_{p^n}$ is also a field, and its field structure picks out a particular ultrametric norm.  It is reasonable to suppose that in fairly generic circumstances, the theory \eno{ONOnceMore} flows to a Wilson-Fisher fixed point---assuming we appropriately tune away relevant operators, in particular the mass deformation.  We exhibited in section~\ref{WILSONIAN} (see especially \eno{rlambda}) the discrete transformations that implement the Wilsonian renormalization group for $\mathbb{Q}_{p^n}$, and we used them to analyze the Wilson-Fisher fixed point in an $\epsilon$ expansion, where $\epsilon = 2s-n$ is the dimension of $\lambda$ in the Gaussian theory.

In section~\ref{VASILIEV}, we employed large-$N$ methods with a Hubbard-Stratonovich field $\sigma$, whose equation of motion sets $\sigma = \phi^i \phi^i$ up to a factor.  Working to the leading non-trivial order in $N$, we obtained our main results so far, the anomalous dimensions
 \eqn{Anomalous}{
  \gamma_\phi &\equiv \Delta_\phi - {n-s \over 2} = \Res_\delta g_\phi(\delta) + 
    {\cal O}(1/N^2)  \cr
  \gamma_\sigma &\equiv \Delta_\sigma - s = \Res_\delta g_\sigma(\delta) +
    {\cal O}(1/N^2) \,,
 }
where $\Res_\delta g(\delta)$ means the residue of a meromorphic function $g(\delta)$ at $\delta=0$.  We found
 \eqn{gResults}{
  g_\phi(\delta) &= {1 \over N} 
    {{\rm B}(n-s,-s+\delta) \over {\rm B}(n-s,n-s)}  \cr
  g_\sigma(\delta) &= -{2 \over N} {{\rm B}(n-s,-s+\delta) \over {\rm B}(n-s,n-s)} + 
     {1 \over N} \left( -1 + 2 {{\rm B}(n-s,n-2s) \over {\rm B}(n-s,n-s)} \right) 
       {{\rm B}(\delta,\delta) \over {\rm B}(n-s,n-s)} \,,
 }
where ${\rm B}$ is the variant of the Euler beta function defined in \eno{GammaBeta}.  An important point is that $g_\phi(\delta)$ has no pole at $\delta=0$ for $\mathbb{Q}_{p^n}$, and so $\gamma_\phi = 0$ in this case.  For $\mathbb{R}^n$, $g_\phi(\delta)$ does have a pole, and one easily recovers standard results \cite{Vasiliev:1981yc,Vasiliev:1981dg} for $\gamma_\phi$ upon setting $s=2$.  On the other hand, $g_\sigma(\delta)$ generically has a pole both for $\mathbb{R}^n$ and for $\mathbb{Q}_{p^n}$.  We have checked that its residue gives $\gamma_\sigma$ in accord with the standard results  for the case of $\mathbb{R}^n$, and in accord with results from the Wilsonian approach of section~\ref{WILSONIAN} for $\mathbb{Q}_{p^n}$.

\subsection{Higher derivative $O(N)$ models on $\mathbb{R}^n$}

In order to compare with standard results in the literature, we have generally set $s=2$ when considering the $O(N)$ model on $\mathbb{R}^n$.  One could reasonably ask, what happens if we lift this restriction?  The large $N$ analysis leading from \eno{ONOnceMore} to \eno{Anomalous}-\eno{gResults} remains valid.  For generic $s$, $g_\phi(\delta)$ has no pole at $\delta=0$.  As for $\mathbb{Q}_{p^n}$, this is associated with having only a finite renormalization of $G_{\phi\phi}$ rather than an anomalous dimension.  For $\mathbb{Q}_{p^n}$ we understand this as a consequence of the non-renormalization argument of \cite{Lerner:1989ty}, following quite generally from ultrametricity.  In $\mathbb{R}^n$ a non-local kinetic term is not expected to be renormalized~\cite{Honkonen:1988fq} due to the fact that Wilsonian renormalization leads to correction terms polynomial in momenta---in other words local derivative couplings in position space which do not affect the non-local kinetic piece. (This reasoning is equally valid in $\mathbb{Q}_{p^n}$, but due to the ultrametricity of the $p$-adic norm a stronger version holds: As discussed in section~\ref{NONRENORMALIZATION}, no derivative couplings are {\it ever} generated.) Theories with a non-local kinetic term in $\mathbb{R}^n$ were already studied in \cite{Fisher:1972zz,Sak:1973a}.  Fisher, Ma and Nickel~\cite{Fisher:1972zz} considered precisely the theory described by \eno{ONOnceMore} in $\mathbb{R}^n$ and computed critical exponents in the $\epsilon$ expansion and at large $N$, in the range $n/2<s<2$. The large $N$ results presented in \eno{Anomalous}-\eno{gResults} find perfect agreement with the anomalous dimensions $\gamma_\phi$ and $\gamma_\sigma$ extracted from the critical exponents:
 \eqnRn{FMNDims}{
  \eta &= 2 -s + 2\gamma_\phi = 2 -s  + \mathcal{O}(1/N^2) \cr
  {1 \over \gamma} &= \left({s - 2\gamma_\phi \over n-s - \gamma_\sigma}\right)^{-1} \!\!= 1 - {2s-n \over s} - {8 \over N}  {\Gamma_{\rm Euler}({s \over 2})^2 \Gamma_{\rm Euler}(n-s) \over s\, \Gamma_{\rm Euler}(s-{n \over 2}) \Gamma_{\rm Euler}({n \over 2}) \Gamma_{\rm Euler}({n-s \over 2})^2}  \cr 
 &  \quad \quad \quad \times \left[ {\Gamma_{\rm Euler}({s \over 2}) \Gamma_{\rm Euler}(n-s) \Gamma_{\rm Euler}({n\over 2}-s) \Gamma_{\rm Euler}({3s-n \over 2}) \over  \Gamma_{\rm Euler}(s) \Gamma_{\rm Euler}(n-{3s \over 2}) \Gamma_{\rm Euler}(s-{n \over 2}) \Gamma_{\rm Euler}({n -s \over 2})} - {1 \over 2} \right] + \mathcal{O}(1/N^2).
 }
In \eno{FMNDims}, $\eta$ and $\gamma$ are critical exponents computed in \cite{Fisher:1972zz}, while $\gamma_\phi$ and $\gamma_\sigma$ are obtained from \eno{Anomalous}-\eno{gResults}.  Generically for $s \geq 2$, the local kinetic term $\sim (\partial \phi)^2$, generated from Wilsonian considerations, becomes more relevant and dominates the non-local kinetic piece, resulting in a non-vanishing anomalous dimension for $\phi$ found by setting $s=2$ in \eno{Anomalous}-\eno{gResults}.  The discontinuity in $\gamma_\phi$ at $s=2$ can be removed by accounting for the competition between the  local kinetic term induced from renormalization and the non-local kinetic piece, with the local kinetic term argued to become more relevant at $s=s_\star < 2$ in such a way that $\gamma_\phi$ is continuous along $s$~\cite{Sak:1973a,Honkonen:1988fq} (see also~\cite{Parisi:2014a,Parisi:2014b,Defenu:2014bea}).\footnote{As this paper was nearing completion, we became aware of forthcoming work by S.~Rychkov and collaborators on related issues.} In this paper, however, we have concerned ourselves with a $(\phi^i \phi^i)^2$ deformation as shown in \eno{ONOnceMore} with all other relevant deformations appropriately tuned away, and the results presented in \eno{Anomalous}-\eno{gResults} are valid as long as that continues to hold.

Perhaps a more interesting question is what happens when $s=4$, or $6$, or some larger even number.  At precisely these values, the original model \eno{ONOnceMore} recovers locality.  It is then a higher derivative version of the $O(N)$ model.  Let's consider the case $s=4$ for the sake of a focused discussion.  Then
 \eqnRn{SBoxBox}{
  S = \int dx \left[ {1 \over 2} (\square \phi^i)^2 + {\lambda \over 4!} (\phi^i \phi^i)^2 \right] \,,
 }
from which it follows that
 \eqnRn{BoxBoxDims}{
  [\phi^i] = {n-4 \over 2} \qquad\qquad [\lambda] = 8-n \,,
 }
and so we see that the upper critical dimension is $n=8$, while the lower critical dimension is $n=4$.  Between the upper and lower critical dimension, the interaction term $(\phi^i \phi^i)^2$ is relevant, so it triggers a renormalization group flow which we may suspect leads to a new critical theory in the infrared---provided relevant deformations are appropriately tuned away.  Precisely this sort of flow was considered in $8-\epsilon$ dimensions in \cite{Hornreich:1975a}, and the infrared critical theory was referred to as a Lifshitz point.  Setting $s=4$ in \eno{Anomalous}-\eno{gResults} leads to the following predictions for the anomalous dimensions at Lifshitz points:
 \eqnRn{HighWF}{
  \begin{tabular}{c||c|c|c}
   $n$ & $5$ & $6$ & $7$ \\ \hline\hline
   $N\gamma_\phi$ & ${48 \over 35\pi^2}$ & 0 & $-{128 \over 315 \pi^2}$ \\[1pt] \hline
   $N\gamma_\sigma$ & $-{1408 \over 105 \pi^2}$ & $-{14 \over 3}$  & $-{15872 \over 315 \pi^2}$
  \end{tabular}
 }
up to ${\cal O}(1/N^2)$ corrections to both $\gamma_\phi$ and $\gamma_\sigma$ in each case.  These results were anticipated in \cite{Hornreich:1975b}; in fact, results were given there for fixed $s=4$ and arbitrary $n \in (4,8)$ in the form
 \eqnRn{HDims}{
  \eta_{\ell 4} &= 4-s+2\gamma_\phi \Big|_{s=4} = {1\over N} {(8-n) \over n(n+2)}{ 3 \times 2^{n-2}  \over \pi^{3/2}} {  \Gamma_{\rm Euler}({n-3 \over 2}) \over   \Gamma_{\rm Euler}({n \over 2})} \sin {\pi n \over 2}  + \mathcal{O}(1/N^2) \cr
  \gamma_\ell &= {s - 2\gamma_\phi \over n-s - \gamma_\sigma}\Bigg|_{s=4} = \left({n\over 4}-1\right)^{-1} - {1 \over N} { \Gamma_{\rm Euler}(n-4) \over \Gamma_{\rm Euler}({n\over 2}) \Gamma_{\rm Euler}({n \over 2} -2)^2 \Gamma_{\rm Euler}(4-{n\over 2}) } \cr 
  & \quad \quad \quad \times \left({n\over 4}-1\right)^{-2} \left[ 1 + {(10-n)(n-5) \over 3} + {3(n-6)(n-8) \over 4(n+2)} \right] + \mathcal{O}(1/N^2).
 }
In \eno{HDims}, $\eta_{\ell 4}$ and $\gamma_\ell$ are quantities defined and computed in \cite{Hornreich:1975b}.  Explicit expression in terms of $\Gamma_{\rm Euler}$ can be derived for $\gamma_\phi$ and $\gamma_\sigma$ starting from \eno{Anomalous}-\eno{gResults} with $s$ set equal to $4$, and when this is done, perfect agreement with \eno{HDims} is found.

The expressions for $\gamma_\phi$ and $\gamma_\sigma$ that we gave in \eno{Anomalous}-\eno{gResults} go smoothly to zero at both the upper and lower critical dimensions.  At the upper critical dimension, the natural expectation is that the only fixed point is the Gaussian theory, and turning on $\lambda$ causes us to run logarithmically away from it.  At the lower critical dimension (namely four), we recover the four-dimensional sigma-model considered in \cite{Gava:1978gd}, and the value given in \cite{Gava:1978gd} for the anomalous dimension of $\phi$ just above the lower critical dimension in an epsilon expansion matches the $s=4$ case of the $1/N$ result \eno{Anomalous}-\eno{gResults}. We comment further on the lower critical dimension at the end of section~\ref{FLOWS}.

\subsection{A bound on the higher derivative action}
\label{BOUND}

To properly understand the field theory \eno{SBoxBox}, we should list the relevant deformations: for $n \geq 6$,
 \eqnRn{Srel}{
  S_{\rm rel} = \int dx \left[ {w \over 2} \phi^i \square \phi^i + {r \over 2} \phi^i \phi^i \right]
    \,,
 }
where $[w] = 2$ and $[r] = 4$.  (Of course, $\phi^i \square \phi^i = -(\partial\phi^i)^2/(2\pi)^2$ up to a total derivative which we can discard.)  With these extra terms added, the action may no longer be everywhere nonnegative, and one might wonder about runaway instabilities.  The aim of this section is to provide an estimate which shows that by adding a suitable constant term to the lagrangian, we can make it once again nonnegative.  This is the sense in which the action is bounded below.  While the estimates we give are fairly trivial, they are worthwhile to see given the prevalence of instabilities and ghosts in higher derivative theories after passing to a Hamiltonian or Lorentzian setting.

For $n<6$, $O(N)$ singlet operators schematically of the form $\phi^2 (\partial\phi)^2$ become relevant as well, and we can proceed to $\phi^4 (\partial\phi)^2$ operators when we have $n<5$.  Such a large assortment of terms would complicate the story too much for us to give simple estimates, so let's stipulate $n \geq 6$ in this section.

We may bring the action into a form considered for example in \cite{Hawking:1985gh,deUrries:1998obu} by trading $w$ and $r$ for two mass parameters, $m_1$ and $m_2$:
 \eqnRn{Stotal}{
  S + S_{\rm rel} = \int dx \left[ {1 \over 2} \phi^i q(\square) \phi^i + 
    {\lambda \over 4!} (\phi^i \phi^i)^2 \right]
 }
where 
 \eqnRn{qForm}{
   q(\square) = (\square + m_1^2) (\square + m_2^2) \,.
 }
We can assume $m_1^2 < m_2^2$ without loss of generality, but we cannot necessarily assume that the $m_i^2$ are positive.  Aficionados of Pauli-Villars regulators will immediately recognize \eno{qForm} and the consequent tree-level momentum space propagator:
 \eqnRn{PVform}{
  G^{(0)}_{\phi\phi}(k) = {1 \over (k^2+m_1^2)(k^2+m_2^2)} = 
   {1 \over m_2^2-m_1^2} \left( {1 \over k^2+m_1^2} - {1 \over k^2+m_2^2} \right) \,.
 }
The Pauli-Villars strategy is to let the $1/k^4$ behavior of this propagator improve UV behavior, and then at the end of a computation take $m_2$ large while $m_1$ remains finite.  (Normally in a Pauli-Villars context one would rescale $\phi$ by a power of $m_2^2-m_1^2$ to get rid of the $1/(m_2^2-m_1^2)$ prefactor in the last expression in \eno{PVform}.)  The minus sign on the $1/(k^2+m_2^2)$ term in \eno{PVform} is understood as an indication of ghosts (i.e.~negative norm states in the Hilbert space) in a canonical quantization approach.  Indeed, pathological features of higher derivative scalar field theories have been explored extensively: see for example \cite{Hawking:1985gh,Jansen:1993jj,deUrries:1998obu,Hawking:2001yt} and references therein.  Typical pathologies hinge on a Hamiltonian construction in which one sees an instability along the lines of Ostrogradsky's theorem~\cite{Ostrogradski:1850a}, and/or failures of reflection positivity \cite{Hawking:1985gh} that lead to negative norm states in a canonical quantization approach.  In a Euclidean quantum field theory setting, these pathologies may prove less significant as long as we do not attempt canonical quantization.  Instead, we should form a Euclidean path integral
 \eqnRn{Zpartition}{
  Z = \int {\cal D} \phi \, e^{-S[\phi] - S_{\rm rel}[\phi]} \,,
 }
and then what matters is that the total action should be bounded below and that it should not have flat or nearly flat directions that prevent convergence.  Boundedness can be demonstrated explicitly, as follows.
 \eqnRn{CSstep}{
  \left| \int dx \, {1 \over 2} (m_1^2+m_2^2) \phi^i \square \phi^i \right| &\leq
   \left( \int dx \, {1 \over 4\xi} (m_1^2+m_2^2)^2 \phi^i \phi^i \right)^{1/2}
   \left( \int dx \, \xi (\square \phi^i)^2 \right)^{1/2}  \cr
  &\leq \int dx \left[ {\xi \over 2} (\square\phi^i)^2 + {1 \over 8\xi} (m_1^2+m_2^2)^2 \phi^i \phi^i \right]
   \,,
 }
where the first inequality is Cauchy-Schwarz and the second is the arithmetic-geometric mean inequality, and $\xi$ is any positive real number.  Plugging \eno{CSstep} into \eno{Stotal}, we arrive at
 \eqnRn{Sbounded}{
  S + S_{\rm rel} &\geq \int dx \Bigg[ {1-\xi \over 2} (\square \phi^i)^2 - 
     {1 \over 8} \left( {1-\xi \over \xi} (m_1^2+m_2^2)^2 + (m_1^2-m_2^2)^2 \right) \phi^i \phi^i  \cr
    &\qquad\qquad{} 
    + {\lambda \over 4!} (\phi^i\phi^i)^2 \Bigg] \,.
 }
We must choose $\xi \in (0,1)$ in order to get the derivative term on the right hand side of \eno{Sbounded} to be positive definite, so as to make the lower bound strong when the $\phi^i$ are highly oscillatory.  Choosing $\xi \in (0,1)$ makes the mass term on the right hand side of \eno{Sbounded} negative, which seems like the beginning of an instability; but as long as $\lambda > 0$ the overall value of the lagrangian density is bounded below.  We could adjust the lagrangian density by a constant term (which is after all a relevant deformation) to achieve an action which can be shown to be nonnegative through the approach outlined in \eno{CSstep}-\eno{Sbounded}.  In short, the situation is no worse than the case of the usual $O(N)$ model on $\mathbb{R}^n$ with negative mass squared.  It should be borne in mind that the inequalities might be far from sharp.  So the actual behavior of $S+S_{\rm rel}$ could be somewhat better than we have demonstrated.

\subsection{Qualitative features of renormalization group flows}
\label{FLOWS}

Starting from the free massless higher derivative theory $S_0 = \int dx \, {1 \over 2} (\square\phi^i)^2$, let's consider what renormalization group flows there must be, indicating in each case what the likeliest outcome is in the infrared.  For simplicity we avoid consideration of deformations which lead to soft or spontaneous breaking of translational or rotational symmetry on $\mathbb{R}^n$.  We assume that $n>4$ so that the dimension of $\phi^i$ is positive, and we assume $n<8$ so that we have relevant deformations, namely $\phi^2$, $(\partial\phi)^2$, or $\phi^4$, where we omit $O(N)$ indices for brevity.  Let's consider in turn the deformations with respect to each:
 \begin{itemize}
  \item Deforming only by $\phi^2$ with a positive coefficient looks boring in the sense that it can only lead to a theory in which there are no light degrees of freedom.  We exclude the case of adding $-\phi^2$ to the action because then there really would be a runaway instability.
  \item Deforming only by $(\partial\phi)^2$---where again to avoid instability we must insist on a positive coefficient---leads trivially to the massless two-derivative Gaussian theory, with action (proportional to) $\int dx \, {1 \over 2} (\partial\phi)^2$.  We say ``trivially'' because there are no loop diagrams.  All we are doing is setting $m_2 \neq 0$ in \eno{qForm} while keeping $m_1^2=0$.  The only non-trivial Green's function is the two-point function $G_{\phi\phi}(k) \propto {1 \over k^2} - {1 \over k^2 + m_2^2}$, the same as for a free massless scalar plus a Pauli-Villars regulator.  Passing to the regime $|k| \ll m_2$ amounts to excising the Pauli-Villars part of the propagator.
  \item Deforming by $(\partial\phi)^2$ and $\phi^4$, with positive coefficients for each, while tuning the coefficient of $\phi^2$, should enable us to again reach massless two-derivative Gaussian theory.  The key point is that $(\partial\phi)^2$ is more relevant than the original $(\square\phi)^2$ term, so the latter drops out; and in the new dimension counting based on $(\partial\phi)^2$, the interaction term $\phi^4$ is irrelevant, so it too should attenuate away as we proceed toward the infrared.  In the process, $\phi^2$ terms are generated, so to wind up at the free massless Gaussian theory rather than a massive theory we must tune $\phi^2$.
  
We could also take the Pauli-Villars point of view and reason that our deformed theory in this case is a Pauli-Villars regularization of the usual two-derivative $O(N)$ model.  Since we are above the upper critical dimension of this two-derivative theory, the transition from the disordered state to the ordered state must be described by mean field theory, i.e.~the massless two-derivative Gaussian theory.
  \item Deforming by $\phi^4$ with a positive coefficient while tuning both $(\partial\phi)^2$ and $\phi^2$ should enable us to reach new conformal theories whose anomalous dimensions for integer $n$ are listed in \eno{HighWF}.  Deforming only by $\phi^4$ doesn't make sense because loop effects will presumably generate $(\partial\phi)^2$ and $\phi^2$.  If we don't tune the $\phi^2$ term, we'll wind up with a massive theory, while if we don't tune the $(\partial\phi)^2$ term we could wind up with the two-derivative Gaussian theory.
  
Below $n=6$, new relevant $O(N)$ singlets appear: the aforementioned $\phi^2 (\partial\phi)^2$ operators.  Their coefficients might also need to be tuned in order to arrive at the new conformal field theories whose existence we are hypothesizing.  Relevant operators of this type may be relatively harmless since their dimensions are always higher than the operator $\phi^4$ which is driving the flow.
 \end{itemize}

\begin{figure}[t]
  \centerline{\includegraphics[width=6.5in]{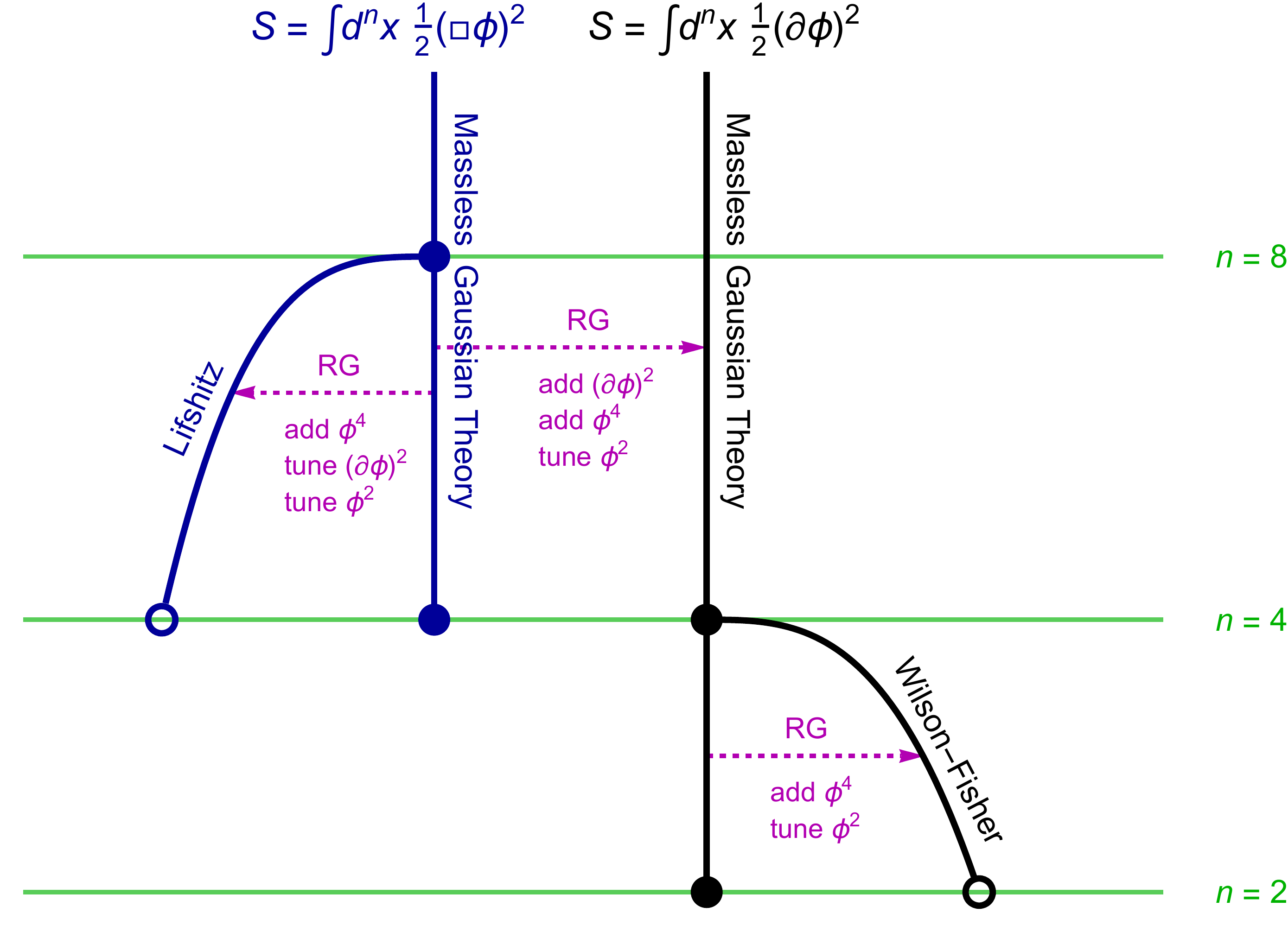}}
  \caption{The four-derivative extension of the space of fixed points of $\phi^4$ theory.}\label{FixedPoints}
 \end{figure}

Altogether, four-derivative $\phi^4$ theory should augment the space of fixed points of the $O(N)$ model as indicated in figure~\ref{FixedPoints}.
If this picture is accepted, the next natural question is what happens at the lower critical dimension.  In the case of two-derivative theories, the key point for $N>1$ is that non-linear sigma models (NL$\sigma$M) on $S^{N-1}$ become renormalizable in $n=2$---though for $N>2$ they are asymptotically free rather than conformal.  In the case $N=1$, the symmetry group is $\mathbb{Z}_2$, and we obtain the $c=1/2$ minimal model as the continuum limit of 2d Ising.  In other words, the NL$\sigma$Ms (or, for $N=1$, the $c=1/2$ minimal model) are at the terminus of the line of Wilson-Fisher fixed points as we proceed downward in dimension.
 
Proceeding by analogy, we might expect in $n=4$ some new way of realizing $O(N)$ symmetry in a renormalizable field theory.  The obvious candidate is a NL$\sigma$M on $S^{N-1}$, where the kinetic term is $(\square \phi^i)^2$ with $\phi^i \phi^i$ constrained to be equal to $1$. Exactly such a theory is considered in \cite{Gava:1978gd}, and the beta-function computed there accords with the natural expectation that for $N$ large enough (larger than $2$) the theory is asymptotically free in the ultraviolet and confining in the infrared. Since the NL$\sigma$M construction is unavailable for $N=1$, we are thrown back on the more abstract proposal that there could be some four-dimensional Euclidean conformal field theory whose natural degrees of freedom we don't know but which realizes a global $\mathbb{Z}_2$ symmetry.

It is of course tempting not to stop with $\phi^4$ theory; as in two-derivative theories one can consider higher powers of $\phi$, leading to new branches of fixed points that fork off the Gaussian theory at dimensions that are successively closer to $n=4$ as one raises the power of $\phi$.  Such fixed points are called multicritical in the two-derivative context because one has to tune several relevant operators to hit the infrared fixed point.  They are thought to connect to minimal models in the lower critical dimension \cite{Zamolodchikov:1986db}.  Multicriticality will be even more pronounced for four-derivative theories, because the list of relevant operators proliferates quickly as we head toward $n=4$ and includes an assortment of two-derivative operators.  Let us nonetheless conjecture that multicritical versions of Lifshitz fixed points above $n=4$ exist for the $O(N)$ model, and that for $N=1$ they are continuously connected with new conformal field theories in $n=4$ which are analogs of unitary minimal models.  These new theories, both in $n=4$ and in higher dimensions, may be amenable to treatment via the conformal bootstrap, similar to \cite{Kos:2015mba}.  If all this is true, then one might hope that other classic field theory constructions in $n=2$ generalize to higher derivative theories in $n=4$; in such a case, we clearly have a lot of work to do to understand what the full picture of four-dimensional Euclidean field theories really comprises!  

One also need not stop with four-derivative theories.  The next case to consider is $\phi \square^3 \phi$ theory.  The upper critical dimension (where $\phi^4$ becomes marginal) is $12$, and the lower critical dimension is $6$.  It is easy to read off from \eno{Anomalous}-\eno{gResults} the anomalous dimensions of $\phi$ and $\phi^2$ at conjectural fixed points anywhere between $n=6$ and $12$.  The list of relevant deformations will be even more extensive than in the four-derivative case, and correspondingly one must expect quite a complex picture of possible renormalization group flows. Problems with canonical quantization and Ostrogradsky instabilities are likely to be ubiquitous in all the higher derivative theories we are considering, but as Euclidean path integral field theories they are probably well defined due to bounds along the lines of section~\ref{BOUND}. In fact, by studying the analytical structure of the conformal blocks of generalized free CFTs (unitary and non-unitary) and nearby Wilson-Fisher critical points, the authors in \cite{Gliozzi:2016ysv,Gliozzi:2017hni} derive expressions for the first terms of the anomalous dimensions of classes of scalar operators in an epsilon expansion, and their results for $\gamma_\phi$ and $\gamma_{\phi^2}$ in theories with (in our notation) $s=n/2$ exactly matches \eno{Anomalous}-\eno{gResults} in arbitrary dimension.

\subsection{A lattice implementation}

Just as ordinary two-derivative $\phi^4$ theory (with real-valued $\phi$, i.e.~$N=1$) is realized as a continuum limit of the Ising model with nearest neighbor interactions, we might expect four-derivative $\phi^4$ theory to be realized as a continuum limit of an Ising model with next-to-nearest neighbor interactions.  We have in mind particularly a lattice action along the lines of the anisotropic next-to-nearest neighbor Ising model (ANNNI for short) \cite{Elliott:1961zz,Fisher:1980zz}, but as isotropic as the underlying lattice allows:
 \eqn{IsingNNN}{
  S = K \sum_{\vec{x} \in \mathbb{Z}^n} (\square\sigma_{\vec{x}})^2 + 
   J \sum_{\vec{x} \in \mathbb{Z}^n} \sum_{\vec{y} \sim \vec{x}} (\sigma_{\vec{x}}-\sigma_{\vec{y}})^2
    \,,
 }
where we define a lattice laplacian
 \eqn{LatticeLap}{
  \square \sigma_{\vec{x}} = \sum_{\vec{y} \sim \vec{x}} (\sigma_{\vec{x}} - \sigma_{\vec{y}}) \,.
 }
The notation $\sum_{\vec{y} \sim \vec{x}}$ means that we hold $\vec{x}$ fixed and sum over all $\vec{y}$ which are nearest neighbors of $\vec{x}$, which is to say $2n$ nearest neighbors when we work on the lattice $\mathbb{Z}^n$.  With the action \eno{IsingNNN} in hand, we can define a partition function
 \eqn{PathIntegral}{
  Z = \sum_\sigma e^{-S} \,,
 }
where the sum is over all possible spin configurations.  If we set $K=0$, then according to standard reasoning, there is a phase transition between ordered and disordered phases that occurs at a special value of $J$, and it will have mean field theory critical exponents when $n>4$ because $n=4$ is the upper critical dimension of two-derivative $\phi^4$ theory.  But if $4 \leq n < 8$, we should be able to find a critical point not described by mean field theory by tuning both $K$ and $J$.  Instead, the critical point should be described by the endpoint of a renormalization group flow from the massless four-derivative Gaussian theory, triggered by $\phi^4$ deformation and with the relevant operators $\phi^2$ and $(\partial\phi)^2$ appropriately tuned---the Lifshitz point of \cite{Hornreich:1975a}.  A caveat, as previously noted, is that as one gets close to the lower critical dimension, additional relevant operators appear, so it is conceivable that more lattice quantities must be tuned than just $K$ and $J$.  For $n\geq 6$ this should not be a problem.

Similar lattice constructions can obviously be given for $N>1$ theories.  We could even construct next-to-next-to-nearest neighbor models which should give Lifshitz-like critical points in dimensions between $6$ and $12$---but the computational difficulties associated with lattices in such large dimensions, not to mention the number of tunings required to suppress relevant directions, seem likely to make anything beyond next-to-nearest neighbor models impractical.  The recent work \cite{Berche:2012td} indicates that $n=5$ lattice simulations of the Ising model on large enough lattice to see scaling behavior are accessible with modern computational methods.  So it should be possible to do a direct search on the lattice for non-mean-field critical behavior in \eno{IsingNNN} in $n=4$, $5$, and maybe $6$.  It would also be interesting to study finite-range Ising models on the Bethe lattice, whose recursive structure often leads to exactly solvable models and whose exponential growth mimics infinite dimension \cite{Eggarter:1974zz,Zittartz:1974zz}.  Such studies might eventually lead us back to the $p$-adics through the holographic relation of the Bethe tree with coordination number $p+1$ to the $p$-adic numbers $\mathbb{Q}_p$ on the boundary.

\section{Discussion}
\label{DISCUSSION}

Our main technical result, summarized in \eno{Anomalous}-\eno{gResults}, is the expression of anomalous dimensions $\gamma_\phi$ and $\gamma_\sigma$ as residues at $\delta=0$ of meromorphic functions $g_\phi(\delta)$ and $g_\sigma(\delta)$ of a quantity $\delta$, understood as a shift in the dimension of the Hubbard-Stratonovich field $\sigma$ that we impose as a regulator and then remove at the end of the calculation.  These meromorphic functions come from diagrammatic amplitudes of the form
 \eqn{GammaD}{
  I_{V,m}(z_a) = \int d^V x \, 
   \left( \prod_{a=1}^m \prod_{i=1}^V {1 \over |x_i-z_a|^{\delta_{ia}}} \right)
   \left( \prod_{i \neq j} {1 \over |x_{ij}|^{\delta_{ij}}} \right)
 }
where $x_{ij} = x_i-x_j$.  $V$ is the number of internal vertices, each at a spatial location $x_i$.  The notation $\int d^V x$ means that we are integrating $x_i$ over all space; and ``space'' here could be $\mathbb{R}^n$ or $\mathbb{Q}_{p^n}$.  The number of external vertices is $m$, each at a spatial location $z_a$.  Based on the Feynman rules for the particular theory under consideration (the $O(N)$ model in our case), we are able to assign values to the exponents $\delta_{ij}$ and $\delta_{ia}$ which are linear functions of the regulator $\delta$.  Then $I_{V,m}(z_a)$ becomes a meromorphic function of $\delta$, and a linear combination of several such functions, each deriving from a different diagram, gives us the meromorphic functions $g_\phi(\delta)$ and $g_\sigma(\delta)$ that we are eventually interested in.

In general, such integrals give complicated answers.  However, there are particular cases where the answer simplifies.  If $m=2$ and $V=1$, then $I_{V,m}$ is just a convolution, so the answer is expressed naturally in terms of the appropriate variant of the beta function together with a power of $z_{12}$.  If $m=3$ and $V=1$, then the same thing happens again provided the exponents obey a sum rule: This is the star-triangle identity \eno{StarTriangle}.  The striking point about $O(N)$ model calculations, to the order we have exhibited here, is that all the amplitudes of interest for the computation of anomalous dimensions are expressible as products of the beta function times power-law functions of the $z_a$.\footnote{Final expressions for the functions $g_\phi(\delta)$ and $g_\sigma(\delta)$ involve factors of ${\rm B}(n-s,n-s)$ in the denominator for a special reason: this particular beta function appears in the leading order propagator for $\sigma$.  In other words, negative powers of ${\rm B}(n-s,n-s)$ appear in the meromorphic functions only because they appear in the coefficients we must use to combine the $I_{V,m}(z_a)$ into 1PI amplitudes.} At the Archimedean place, the question of which transcendentals appear in anomalous dimensions at various loop orders is well-studied, and it is known that at order $\mathcal{O}(1/N^3)$ transcendentals that cannot be obtained by differentiation of zeta-functions begin to make an appearance \cite{Broadhurst:1996ur}.

There is an interesting connection between the amplitudes $I_{V,m}$ and string scattering amplitudes.  The beta function appears in the star-triangle identity precisely as it appears in four-point scattering amplitudes of tachyons, i.e.~the Veneziano or Virasoro-Shapiro amplitude.\footnote{We should bear in mind that our ${\rm B}_{\mathbb{R}}(t_1,t_2)$ is not the same as ${\rm B}_{\rm Euler}(t_1,t_2)$; rather, ${\rm B}_{\mathbb{R}}(t_1,t_2)$ is a crossing-symmetric combination of Euler beta functions.  Thus when we refer to the Veneziano amplitude, we really mean the crossing-symmetric combination without Chan-Paton factors.}  The star-triangle identity is in fact a generalization of the way one obtains the Veneziano amplitude by integration over the position of one vertex operator over the boundary of the string.  The sum rule on the exponents is understood in this context as related to momentum conservation plus the on-shell condition for external string states.  Generalizations of the Veneziano amplitude to integrations over all of $\mathbb{R}^n$ were considered in \cite{Brower:1971nd}, while generalizations to integrations over $\mathbb{Q}_p$ are the foundation of $p$-adic string theory \cite{Freund:1987kt,Freund:1987ck,Brekke:1988dg}.  If we add more internal vertices, then in the string theory context, instead of a four-point scattering amplitude, we would be considering a higher point amplitude---still at tree level.  Might we understand the expression of the $I_{V,m}$ integrals we need for anomalous dimensions in the $O(N)$ models in terms of products of beta functions as a consequence of a factorization property of string amplitudes?

The analogy between diagrammatic amplitudes in scalar field theory and string scattering helps our intuition in understanding why the anomalous dimension $\gamma_\phi$ vanishes for the $O(N)$ model defined over $\mathbb{Q}_{p^n}$, but not for the usual $O(N)$ model defined over $\mathbb{R}^n$.  We saw in section~\ref{SELFONE} that after canceling a quadratic divergence with local counterterms, the amplitude in the $p$-adic case had no further divergences, but in the Archimedean case a logarithm appeared that gave rise to the anomalous dimension.  From the point of view of meromorphic functions, we wound up with an integral $I_D$ that located us at a pole in the Archimedean case which would be understood in string scattering terms as an infinitely sharp resonance due to the exchange of a first-excited string state (where tachyons are counted as the unexcited state).  The absence of a pole in the $p$-adic case implies the vanishing of $\gamma_\phi$ and corresponds to the fact that the $p$-adic string has only one state in its spectrum, namely the tachyon.  In general we would like to associate divergences in field theory with on-shell divergences in string scattering amplitudes.

Once we express diagrammatic amplitudes in the form \eno{GammaD}, it is natural to consider a large generalization, in which we replace $\mathbb{R}^n$ or $\mathbb{Q}_{p^n}$ with some homogeneous space---not necessarily Archimedean.  The propagators $1/|x|^{2\Delta}$ would then naturally be replaced by representations of a group which fixes a point in the homogeneous space.\footnote{The group of interest is generally {\it not} the full group preserving a point.  For instance, in the case of $\mathbb{R}^n$ equipped only with conformal structure, special conformal transformations are excluded even though they preserve the origin.  If $G$ admits an Iwasawa decomposition $G=KAN$, and $M$ is the subgroup of $K$ comprising elements which commute with all of $A$, then on the homogeneous space $G/MAN$ the generalized propagators would be representations of $M$ and $A$.  See the related discussion in \cite{Gadde:2017sjg}.}  For $\mathbb{R}^n$ equipped only with conformal structure rather than full metrical structure, this group would consist of dilations and rotations around the origin, so on top of $1/|x|^{2\Delta}$ we could get a factor depending only on the direction of $x$ and providing a unitary representation of the rotation group.  In common parlance, we could consider operators with spin.  For $\mathbb{Q}_{p^n}$, the natural notion replacing spin hinges on multiplicative characters, as remarked in \cite{Heydeman:2016ldy}; so in place of $1/|x|^{2\Delta}$ we would have $\theta(\hat{x})/|x|^{2\Delta}$ where $\hat{x} \equiv x|x|$ is a unit in $\mathbb{Q}_{p^n}$ and $\theta$ is a unitary multiplicative character of the group of units.  It seems likely that there are significant generalizations of the beta functions we have used, related to convolving generalized propagators.  Star-triangle identities and more general diagrammatic amplitudes may be similarly capable of generalization, and the important question becomes what kind of meromorphic functions appear and how their poles translate into anomalous dimensions, or appropriate generalizations thereof.  It would be fascinating to try to extend standard quantum field theoretic notions of locality and renormalizability to this more general setting.

\clearpage

\section*{Acknowledgments}

We thank P.~Witaszczyk for collaborations on the early stages of this project.  We have benefited from discussions with C.~Callan, D.~Huse, S.~Sondhi, B.~Stoica, and G.~Tarnopolsky.  This work is supported in part by the Department of Energy under Grant No.~DE-FG02-91ER40671.

\bibliographystyle{ssg}
\bibliography{ONmodel}

\end{document}